\documentclass[fleqn,usenatbib]{mnras}

\usepackage{newtxtext,newtxmath}

\usepackage[T1]{fontenc}

\DeclareRobustCommand{\VAN}[3]{#2}
\let\VANthebibliography\thebibliography
\def\thebibliography{\DeclareRobustCommand{\VAN}[3]{##3}\VANthebibliography}

\usepackage{graphicx}	
\usepackage{amsmath}	

\def \arcsec {$^{\prime\prime}$}
\def \arcmin {$^\prime$}
\def \um {$\upmu$m}

\title[Magnetic Fields and Feedback in G34.26+0.15]{Magnetic Fields Under Feedback: A Case Study of the Massive Star-Forming Hub G34.26+0.15}

\author[Z. Khan et al.]{
Zacariyya Afzal Khan$^{1}$\thanks{E-mail: zacariyya.khan.22@ucl.ac.uk},
Kate Pattle$^{1}$, Sarah F. Graves$^{2}$
\\
$^{1}$Department of Physics \& Astronomy, University College London, Gower Street, London, WC1E 6BT, UK\\
$^{2}$East Asian Observatory, 660 N. A’ohōkū Place, University Park, Hilo, HI 96720, USA\\
}

\date{Accepted 2024 October 07. Received 2024 October 04; in original form 2024 May 29}

\pubyear{2024}
 
\begin{document}
\label{firstpage}
\pagerange{\pageref{firstpage}--\pageref{lastpage}}
\maketitle

\begin{abstract} 
We present 850\um\ polarized observations of the molecular cloud G34.26+0.15 taken using the POL-2 polarimeter mounted on the James Clerk Maxwell Telescope (JCMT). G34.26+0.15 is a hub-filament system with ongoing high-mass star-formation, containing multiple H\textsc{ii} regions. We extend the Histogram of Relative Orientations technique with an alternative application that considers the alignment of the magnetic field to the filaments and a H\textsc{ii} region boundary, denoted as the filament alignment factor ($\xi_{F}$) and the ellipse alignment factor ($\xi_{E}$) respectively. Using these metrics, we find that, although in general the magnetic field aligns parallel to the filamentary structure within the system in the north-west, the magnetic field structure of G34.26+0.15 has been radically reshaped by the expansion of an evolved H\textsc{ii} region in the south-east, which itself may have triggered further high-mass star formation in the cloud. Thus, we suggest high-mass star formation is both occurring through mass accretion as per the hub-filament model from one side, and through compression of gas under stellar feedback from the other. We also use HARP observations of C$^{18}$O from the CHIMPS survey to estimate the magnetic field strength across the cloud, finding strengths of $\sim$0.5-1.4 mG.
\end{abstract}




\section{Introduction}

Magnetic fields have been long been understood to be ubiquitous across the interstellar medium (ISM, \citealt{crutcher2004}), present on circumstellar scales \citep{arcetord2020} through to galactic scales \citep{planck2016}. Given their prevalence and increasing strength at high densities \citep{crutcher2010}, it is unsurprising that magnetic fields have been hypothesized to have an active role in star formation, such as affecting the formation of filamentary structures \citep{soler2013} that can later fragment into pre-stellar cores \citep{wardthomps1994}. However, measurement and mapping of magnetic fields is observationally expensive, and as such large-scale mapping has only been widely achieved relatively recently -- for example, the large-scale mapping of the magnetic fields across the Galactic plane by the Planck satellite \citep{planck2016}, and the development of polarimetric capabilities for sub-millimetre telescopes such as the James Clerk Maxwell Telescope (JCMT, \citealt{friberg2016}) and the Atacama Large Millimetre Array (ALMA, \citealt{nagai2016}). Large-scale surveys on the molecular cloud scale are still ongoing (e.g. the BISTRO survey, \citealt{wardthomps2017}), but our understanding of the exact role of magnetic fields in star formation is still limited, and thus our understanding of star formation as a whole \citep{pattle2023}.

Understanding the role of stellar feedback is also critical to understanding the formation of both low- and high-mass stars in molecular clouds (e.g. \citealt{krumholz2019}). Feedback effects -- such as outflows, stellar winds and supernovae -- allow stars at different life stages to reshape the gas and dust surrounding them, and in many cases can both inhibit and stimulate future star formation, thus likely playing a major role in setting star formation efficiency (SFE, \citealt{krumholz2019}). This is particularly important when considering high-mass star formation, with high-mass stars emitting ionizing radiation even whilst forming, creating H\textsc{ii} regions surrounding them. Given that the process by which high-mass stars are able to accrete to their higher mass after reaching the Eddington luminosity is poorly understood \citep{motte2018}, and the propensity of massive stars to form in clusters, considering how H\textsc{ii} regions reshape their surroundings is key to forming a complete image of the star formation process.

Magnetic fields likely also play some role in high-mass star formation, although whether this role is broadly inhibiting or stimulative is unclear: for example, magnetic fields have been shown to act against global gravitational collapse to inhibit formation \citep{mckee2003}, but also to constrain the rate of H\textsc{ii} expansion across their lifetime \citep{cunningham2018}. At very high densities, as seen in molecular clouds in later stages of high-mass star formation, gravitational effects dominate over magnetic fields. As magnetic fields are still coupled to the ionized gas that is acting under gravity in this scenario, they instead act as a rough tracer of gas flow and structure.

G34.26+0.15 is a massive molecular cloud at a distance of $\sim$3.3 kpc \citep{kuchar1994}, with a central area that is sufficiently bright in the sub-millimetre to have been used as a calibrator source for the JCMT. The central region has been extensively studied due to the multiple UCH\textsc{ii} regions contained within it, as noted by \citet{reid1985}, which are visible in 21cm VLA observations.

These UCH\textsc{ii} regions are  referred to as A, B and C \citep{reid1985}. A and B are extremely compact, being roughly spherical and having radii on the order of $\sim$1000 AU \citep{avalos2009}, and are likely the youngest H\textsc{ii} regions in G34.26+0.15. Region A has been determined to be driven by a B0 Zero-Age Main-Sequence Star (ZAMS, \citealt{campbell2000}), with Region B likely containing the same \citep{avalos2009}. Region C, however, is a cometary UCH\textsc{ii} region, with a bright `head' that contains the majority of the flux from G34.26+0.15's centre, and an optically thin `tail' that extends out roughly 8\arcsec\ (0.3 pc) \citep{reid1985} away from Regions A and B. Although the nature of the tail is debated \citep{gaume1994}, the head likely contains a ZAMS of class O6-O6.5 (\citealt{wood1989}, \citealt{campbell2000}). As such, there is strong evidence that the centre of G34.26+0.15 contains multiple massive stars, in different stages of formation.

Additionally, close to the central mass is a fourth major H\textsc{ii} region, referred to as Region D (\citealt{reid1985}, \citealt{gaume1994}). This appears older and is more expanded than Regions A, B and C, spanning a full parsec in plane-of-sky width. Region D has been suggested to have induced sequential star formation \citep{liu2013} in the wider G34.26+0.15 cloud -- i.e. the shock front generated by its expansion may have compressed molecular gas sufficiently to create densities high enough for further star formation to occur. This possibility will be explored in this paper.

However, despite the detail with which the central area has been observed, the extended structure of G34.26+0.15 has been less well-studied. G34+26+0.15 contains multiple prominent filaments, which converge on the bright central area. This is typical of a hub-filament system, an arrangement proposed by \citet{myers2009}, which is hypothesized to facilitate the formation of massive stars by placing them at the bottom of a large global gravitational potential. Some of this structure has additionally been postulated to be part of the long infrared-dark cloud (IRDC) associated with the star-forming region G34.43+0.24 to the north of G34.26+0.15 \citep{xu2016}.

 The structure of this paper is as follows: Section \ref{sec:obs} of this paper details sub-millimeter polarization observations of G34.26+0.15 made with the James Clerk Maxwell Telescope (JCMT), and the data reduction process undertaken. Section \ref{sec:results} presents our results, including the magnetic field structure of G34.26+0.15 (Section \ref{sec:bfields}) with analysis of this structure, a determination of the filamentary structure of G34.26+0.15 (Section \ref{sec:fils}), and a brief analysis of velocity data (Section \ref{sec:velocity}). We then discuss the implication of these results on the role of feedback in the region in Section \ref{sec:discussion}, with a short discussion of H\textsc{ii} region D. Finally, we summarize our findings in Section \ref{sec:conc}.

\section{Observations}
\label{sec:obs}

The region G34.26+0.15 was observed using the POL-2 polarimeter \citep{friberg2016} on the Submillimetre Common-User Bolometer Array 2 (SCUBA-2) camera \citep{holland2013}, mounted on the James Clerk Maxwell Telescope (JCMT). The JCMT is a single-dish submillimetre telescope with a 15-metre primary reflector, located on Maunakea, Hawaii, at an altitude of 4210 metres.

For these observations, SCUBA-2 operated in the standard POL-2 DAISY scanning mode, intended to optimize the integration time on the central region of the scanning area. However, given the brightness of the region, we also have a good detection of the extended structure out to the edges of the mapped region.

Observations were made at wavelengths of 450\um\ and 850\um\ - for the JCMT, this corresponds to 9.8\arcsec\ and 14.6\arcsec\ effective beam widths, respectively \citep{dempsey2013}. The region was observed for $\sim$6 hours in total, over 12 observations -- these were taken on 29 July 2016, 7-9 June 2017, and 1 August 2017 (Project ID: M16AD003) The observations have central coordinates of RA: 18:53:19 and Declination: +01:15:20.

\begin{figure*}
\begin{minipage}[th]{\textwidth}
    \includegraphics[width=\textwidth]{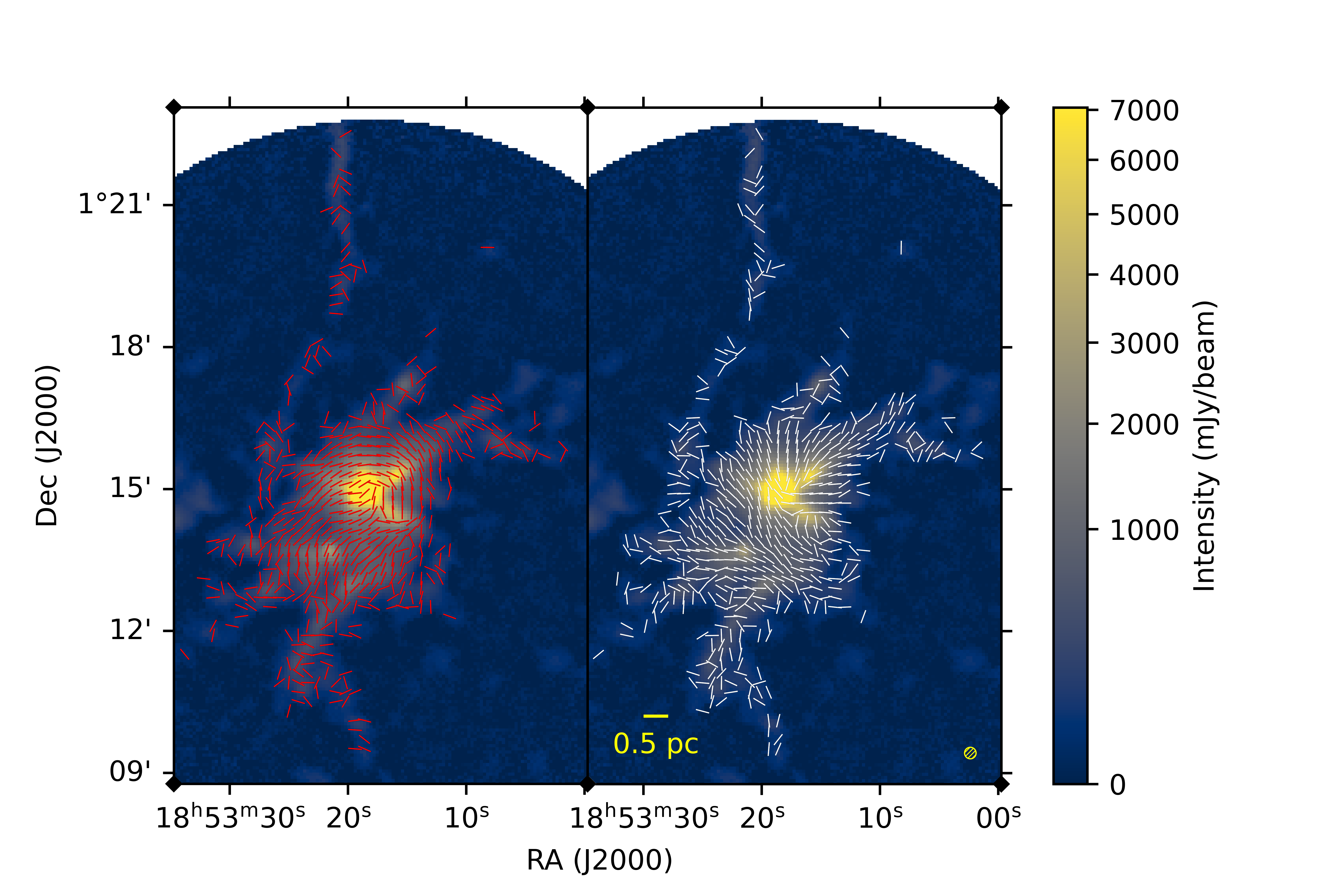}
    \vspace{-10pt}
    \caption{850\um\ Stokes I intensity of G34.26+0.15. Left: The polarization half-vectors are plotted in red, where ${I/\sigma_I} > 5$ and ${P/\sigma_P} > 3$. Right: The inferred magnetic field vectors are plotted in white, where ${I/\sigma_I} > 5$ and ${P/\sigma_P} > 3$. For clarity, all vectors are binned to 3x3 pixels and have constant length regardless of polarization fraction. The JCMT beam size is shown in the lower right of the plot.}
    \label{fig:pol2}
\end{minipage}
\end{figure*}

\subsection{Data Reduction}
\label{sec:datared}

To reduce the POL-2 data, we used the \textit{pol2map} routine included as part of the \textit{Starlink} software package \citep{currie2014}. This routine initially generates the Stokes I, Q and U time streams from the raw 450\um\ and 850\um\ bolometer data, separating out the results into Q, U and I timestreams with the \textit{calcqu} command.

Tpyically, POL-2 data reduction is a two-step process. The first step uses the \textit{makemap} command to generate a Stokes I map only, from the Stokes I timestream only. This uses an iterative algorithm that isolates and removes atmospheric signal in the data (see \citealt{chapin2013} for more details). The second step again uses the \textit{makemap} command and repeats the process for all three Stokes parameters, using the Stokes I map produced from the first cycle as the basis of a mask. In order to ensure the reduced data corresponds to real signal, the initial iterations mark regions with an S/N of greater of than 5 as `real', then iterate out to an S/N of 3. This results in any regions without pixels with a S/N greater than 5 being discarded.

In this cycle, the parameters \textit{mapvar} and \textit{skyloop} are enabled. The \textit{mapvar} parameter takes the variance in each pixel to be taken as the variance of the pixel values across the 12 observations, rather than the propagated error of the combination of bolometer readings. The \textit{skyloop} parameter requires each iteration of the mapmaker used to be performed across all 12 observations at once, rather than iterating each observation individually before combining (\citealt{chapin2013}). See \citet{pattle2021} for a detailed description of the POL-2 data reduction process.

For the reduction of G34.26+0.15, we used the default pixel size of 4\arcsec. The final Stokes maps generated in the second cycle were scaled by a flux conversion factor -- for 850\um\, the standard value is 675 Jy beam$^{-1}$ pW$^{-1}$, as determined by \citet{mairs2021}.

\subsection{Polarization Quantities}
\label{sec:proc}

The data reduction described above produced three maps in 850\um\ -- Stokes I, Q and U -- and the associated variance maps for each. The polarized intensity is given by the equation:

\begin{equation}
    PI=\sqrt{Q^2 + U^2}
	\label{eq:pol_int}
\end{equation}

However, this equation has a systematic bias, producing higher than expected values for polarized intensity (especially at low S/N), as described by \citet{wardle1974}. As such, it is useful to `de-bias' the data set, in order to reduce the effects of this systematic error. To do this, we used the standard modified form of polarized intensity \cite{wardle1974}:

\begin{equation}
    PI^{\prime} = \sqrt{(Q^2 + U^2) - 0.5(\sigma_Q^2 + \sigma_U^2)}
	\label{eq:pol_int_deb}
\end{equation}

where $\sigma_Q$ and $\sigma_U$ are the variances on Q and U, respectively. The uncertainty on both PI and PI\arcmin\ is given by:

\begin{equation}
    \sigma_{PI} = \sqrt{\frac{Q^2\sigma_Q^2 + U^2\sigma_U^2}{Q^2 + U^2}}
	\label{eq:pol_sigma}
\end{equation}

Polarization fraction is taken to be:

\begin{equation}
    P = \frac{PI}{I}
	\label{eq:pol_frac}
\end{equation}

where $I$ is the Stoke I intensity. Again using standard error propagation:

\begin{equation}
    \sigma_{P} = \frac{\sqrt{Q^2\sigma_Q^2 + U^2\sigma_U^2}}{Q^2 + U^2}
	\label{eq:pol_frac_sigma}
\end{equation}

Using these, we apply S/N cuts to the POL-2 data, retaining data where ${I/\sigma_I} > 5$ and where ${P/\sigma_P} > 3$. Additionally, we exclude pixels at the edge of the field, given that signal here is unlikely to be real, being significantly beyond the typical 3\arcmin\ POL-2 observing field. We calculate the polarization angle using the equation:

\begin{equation}
    \theta=\frac{1}{2}\arctan\left(\frac{U}{Q}\right)
	\label{eq:pol_angle}
\end{equation}

Note that $\theta$ is not a true vector angle; instead, it represents a `half-vector', defined only between 0$^{\circ}$ and 180$^{\circ}$.

\section{Results}
\label{sec:results}

\subsection{Magnetic Field Vectors}
\label{sec:bfields}

\begin{figure*}
\begin{minipage}[th]{\textwidth}
    \vspace{-40pt}
    \includegraphics[width=\textwidth]{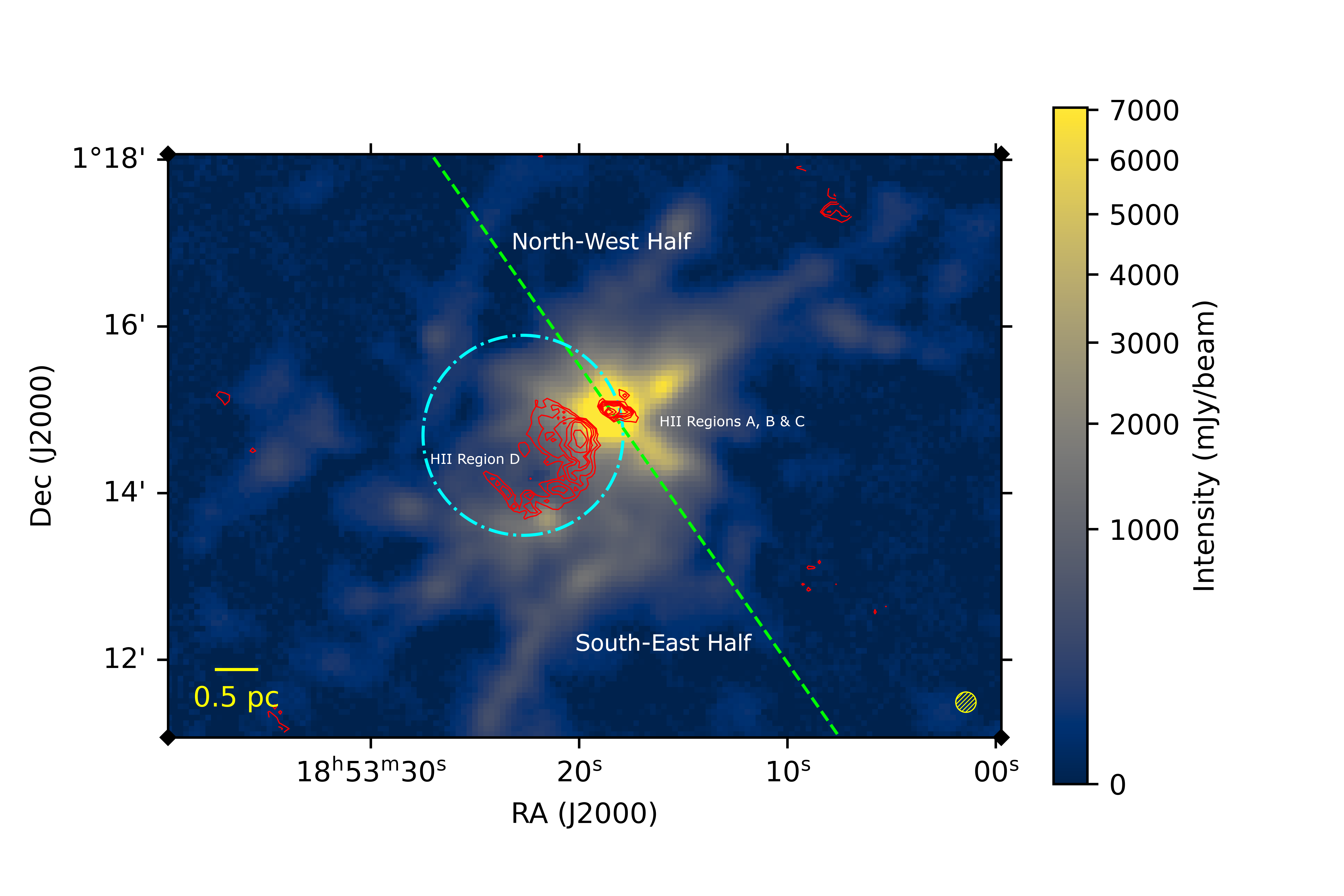}
    \vspace{-30pt}
    \caption{850\um\ Stokes I intensity of G34.26+0.15 detailing the relevant sections used in this paper. Red contours are VLA C-band intensity \citep{crossley2007} at 10\%, 20\%, 40\%,80\% and 100\% levels, used to indicate the positions of major H\textsc{ii} regions in the center of G34.26+0.15. As discussed in Section \ref{sec:bfields}, G34.26+0.15 is split in half for analysis, based on the position of the H\textsc{ii} regions and magnetic contours. The cyan circle represents the bubble from the \citet{simpson2012} catalog, as discussed in Section \ref{sec:feedback}}
    \label{fig:g34_map}
\end{minipage}
\end{figure*}

Using Equation \ref{eq:pol_angle}, the polarization angle was calculated for every pixel across the Stokes I data (subject to the selection criteria described in Section \ref{sec:proc}). As the region is being observed in the sub-millimetre regime, the polarized emission can be assumed to arise entirely from dust -- thus, we can use the polarization angle as a proxy for the preferential orientation of elongated dust grains in the region, with their emission being preferentially polarized along their major axis \citep{andersson2015}. In an extended molecular cloud such as G34.26+0.15, it can assumed that dust grains have been aligned to the local magnetic fields by radiative alignment torque (RAT, \citealt{andersson2015}). As such, we can take the local plane-of-sky magnetic field direction to be perpendicular to the angle of polarized emission.

The inferred magnetic field vectors are shown in Figure \ref{fig:pol2}. Detection of polarization is particularly strong within the central 3 arc-minutes, showing significant organized large-scale structure. As shown in Figure \ref{fig:g34_map}, the region can be broadly split into two halves: the north-west and south-east halves of G34.26+0.15.

The north-west half displays a largely radial magnetic field structure, with magnetic vectors aligned along the similarly radial intensity gradient. In general, this is typical of a hub-filament system \citep{myers2009}: in this scenario, it is assumed that gas and dust are moving towards the core under gravitational influence, with frozen-in magnetic fields tracing this motion. However, as previously noted, there exists a 180 degree ambiguity in the magnetic field directions: the same structure would be observed if the gas were moving radially outwards from the core.


The south-east half displays a markedly different structure -- close to the central hub, the magnetic field runs close-to-perpendicular to the intensity gradient. This almost linear arrangement intersects with what appears to be a distinct substructure, coinciding with a local intensity peak to the south-east of the central hub of G34.26+0.15. This south-eastern `wing' of G34.26+0.15 (more easily identified kinematically, as seen in Section \ref{sec:velocity}) is itself orientated close-to-perpendicular to the central hub, with its own distinct filaments, as discussed in Section \ref{sec:fils}. The magnetic field in the south-east half appears to have been reshaped by the presence of the extended H\textsc{ii} region D, as is discussed later in Section \ref{sec:feedback}. 


\subsection{Alignment Comparison Techniques}
\label{sec:HROs}

\begin{figure*}
\begin{minipage}[th]{\textwidth}
    \centering
    \vspace{0pt}
    \begin{minipage}[t]{0.49\textwidth}
        \includegraphics[width=\textwidth]{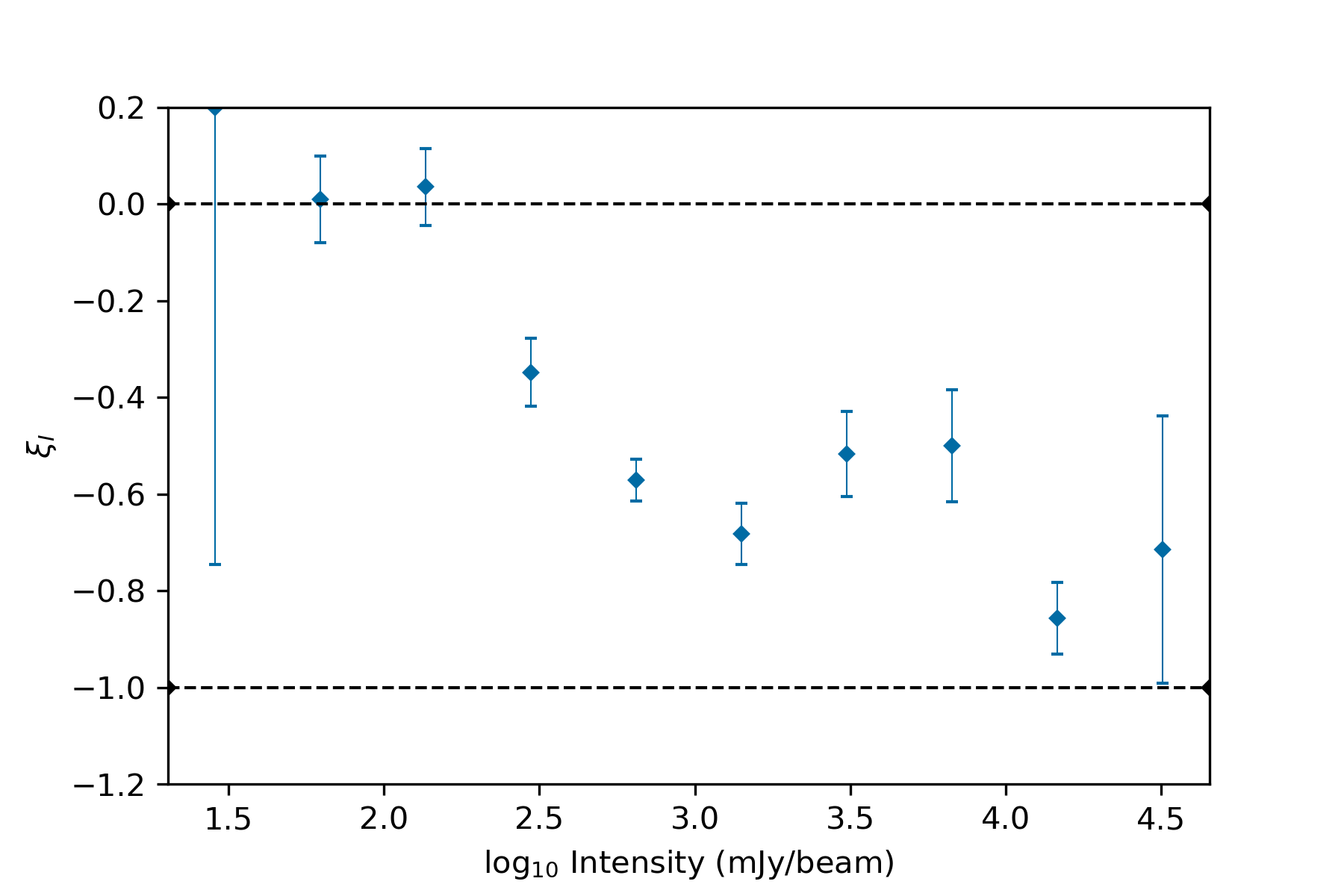}
        \vspace{0pt}
    \end{minipage}
    \begin{minipage}[t]{0.49\textwidth}
        \includegraphics[width=\textwidth]{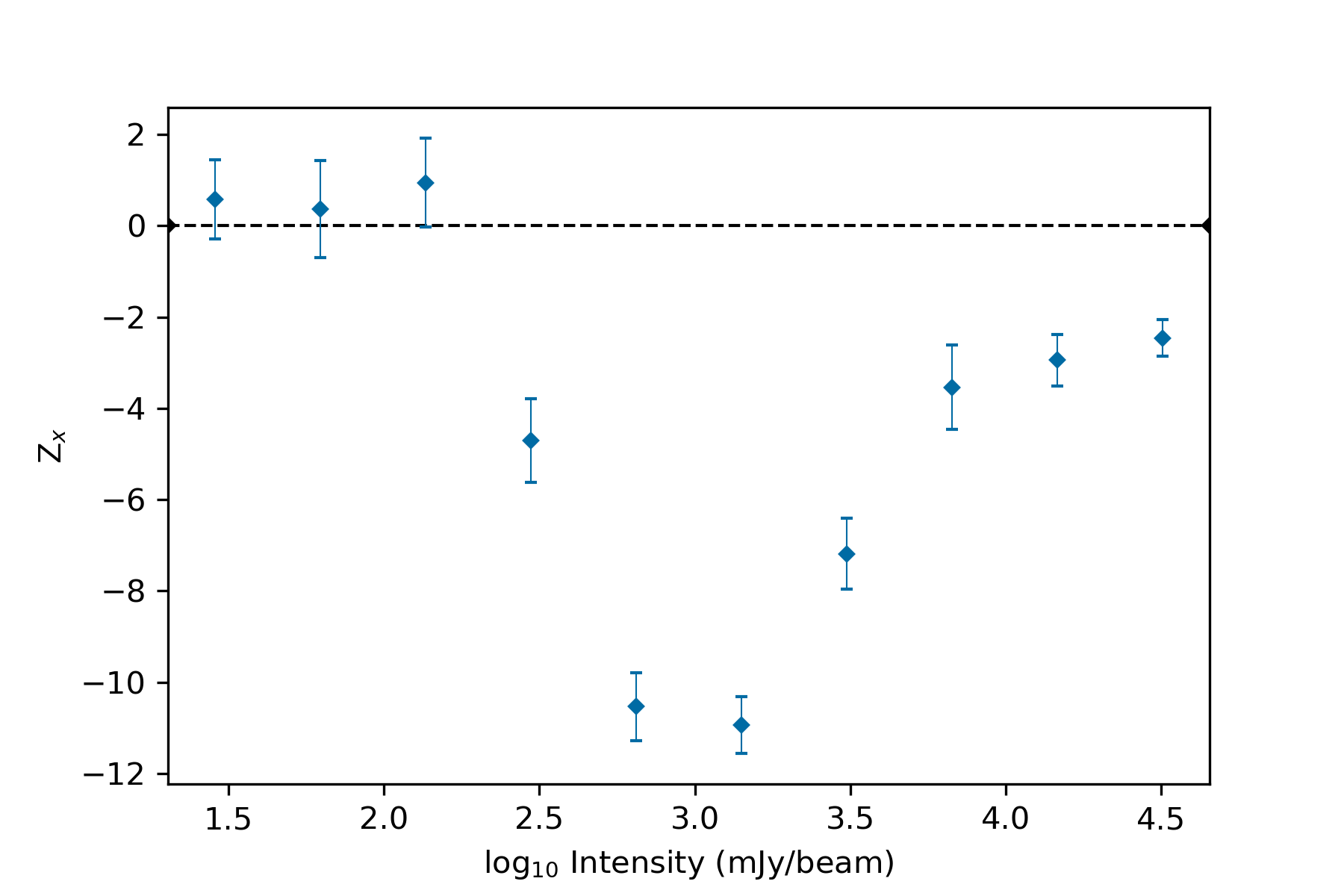}
        \vspace{0pt}
    \end{minipage}
    \caption{Left: HRO showing the shape factor $\xi_{I}$ for 850\um\ Stokes I intensity bins for the G034.26+0.15 field. Errors are given as the combination of the standard errors of $A_e$ and $A_c$, as detailed in Equation \ref{eq:shape_err}. The dotted lines show $\xi_{I}=-1$ and $\xi_{I}=0$. Right: The PRS for each of the Stokes I intensity bins, with errors given as defined in Equation \ref{eq:prs_err}.}
    \label{fig:HRO_full}
\end{minipage}
\end{figure*}

A useful diagnostic for analysing the structure of the magnetic field is the Histogram of Relative Orientations (HRO, \citealt{soler2013}). To construct an HRO, we require an understanding of the density gradients across the region. We use the 850\um\ intensity as a proxy for column density $\Sigma$ for this purpose -- although this assumes isothermality of the dust, and that the dust-to-gas ratio is constant across the region, this is a reasonable approximation \citep{hildebrand1983}. The intensity map was smoothed using 1D Gaussian filters ($\sigma = 3$ pixels, approximately JCMT beam size) across the x and y directions, and the derivative found, producing $\partial\Sigma/\partial x$ and $\partial\Sigma/\partial y$ respectively -- this reduces the effect of noise in the data on individual vectors. These were then combined using the equation:

\begin{equation}
    \psi = \arctan\left(\frac{\partial\Sigma/\partial x}{\partial\Sigma/\partial y}\right),
	\label{eq:gauss_angle}
\end{equation}

This gives the direction of isodensity -- the direction in which the density changes the least across each pixel. For the purposes of creating a HRO, we require the direction of highest density change, which is directly perpendicular to this result. The difference between this direction of gradient $\nabla \Sigma$, and the inferred POS magnetic field $\textbf{B}$ can be expressed as

\begin{equation}
    \phi = \frac{\textbf{B} \times \nabla \Sigma}{\textbf{B} \cdot \nabla \Sigma}.
	\label{eq:phi}
\end{equation}

The HRO technique allows the alignment of $\nabla\Sigma$ and $\textbf{B}$ to be considered, through a metric known as the shape factor. The normalized form of the shape factor \citep{planck2016} is defined as

\begin{equation}
    \xi = \frac{A_c - A_e}{A_c + A_e}.
	\label{eq:shape_factor}
\end{equation}

$A_c$ is defined as the number of parallel alignments within a population (i.e. where $\phi<22.5^\circ$), and $A_e$ is defined as the number of perpendicular alignments (i.e. where $\phi>67.5^\circ$). Values of $\phi$ that do not fit this criterion are not used in this calculation.

The uncertainty of $\xi$, as determined through standard error propagation, is:

\begin{equation}
    \sigma_{\xi}^2 = \frac{2(A_{e}^2 + A_{c}^2)( \sigma_{A_e}^2 + \sigma_{A_c}^2)}{(A_c + A_e)^4}
	\label{eq:shape_err}
\end{equation}

Here, $\sigma_{A_e}^2$ and $\sigma_{A_c}^2$ represent the variances within the bins of $A_e$ and $A_c$ \citep{soler2013}. Constructing a HRO involves splitting the data set into distinct bins (typically intensity or density), and calculating a shape factor for each bin.

An alternative to the HRO technique is the Projected Rayleigh Statistic (PRS, \citealt{jow2017}). Similar to the shape factor described in Equation \ref{eq:shape_factor}, this is applied to a population of $\phi$, but instead uses the equation

\begin{equation}
    Z_x = \frac{\sum_{i}^{n}{\cos{(2\phi_i)}}}{\sqrt{n/2}}.
	\label{eq:prs}
\end{equation}

This statistic is not normalized -- a more positive $Z_x$ suggests perpendicular alignment, and a more negative $Z_x$ suggests parallel alignment of the magnetic field to the density gradient. Unlike the shape factor, the PRS does not omit any values of $\phi$, and thus is able to detect alignments that are not strongly parallel or perpendicular. As described by \citet{jow2017}, the error on the PRS can be estimated as

\begin{equation}
    \sigma_{Z_x}^2 = \frac{2\sum_{i}^{n}\cos{(2\phi_i)}^2-(Z_x)^2}{n}.
	\label{eq:prs_err}
\end{equation}

Figure \ref{fig:HRO_full} shows both the HRO and the PRS for the full POL-2 observation of G34.26+0.15. We use intensity as a proxy for column density: we refer to shape factors measured through comparison of B-field direction to intensity gradients as $\xi_I$, to avoid confusion with alternative shape parameters which are discussed below. In both cases, the field shows no preferential alignment at low intensities, with a preference for parallel alignments as intensity increases. If intensity is taken as a proxy for proximity to the central hub (this is not entirely accurate -- see Section \ref{sec:fils}), this suggests the field is collectively unaligned in the outermost regions, becoming more preferentially parallel to the density gradients towards the center.

However, the two techniques produce significantly different results above an intensity of $10^3$ mJy/beam. $\xi_I$ continues to decreases, indicating increasing parallel alignment towards the center, presenting a minimum of close to $\xi_I=-1$ at above $10^4$ mJy/beam. The $Z(x)$ value, on the other hand, instead actually increases slightly, plateauing at a value just below zero.

Given that these higher intensities (i.e. above $\sim$1500 mJy/beam) mostly correspond to the central bright hub of G34.26+0.15, the discrepancy between $Z(x)$ and $\xi_I$ is likely due to organised magnetic field structure that does not directly align to density gradients. In particular, as discussed in Section \ref{sec:bfields}, the field appears to have been significantly altered by the presence of Region D, appearing almost linear in one half of the core as opposed to the radial arrangement in the other. The PRS may be tracing this more linear structure, while the HRO is only able to trace the parallel components due to not considering non-parallel or -perpendicular alignments. It is also worth noting that H\textsc{ii} regions A, B and C are all located in this high-intensity region within a single JCMT beam, and thus it is likely that the magnetic field structure in the vicinity of these regions is more complex than can be resolved by POL-2.

\subsection{Filament Structure / Analysis}
\label{sec:fils}

As described in Section \ref{sec:fils}, G34.26+0.15 is a complex region with clear filamentary structure -- specifically, it appears to be composed of a central bright hub with filaments branching outwards. Filaments are often modelled as straight structures \citep{pillai2020}, however, given the extent and distinct curved shapes of the filaments seen in G34.26+0.15, this approximation would likely not be adequate for this field.

In order to define this filaments for further analysis, the python module \textit{fil$\_$finder} \citep{koch2015} was used. This creates a mask over a given data set, attempting to fit to filamentary structure with a user-defined size (in this case, filaments with a characteristic width of 0.2 parsecs) along intensity gradients. The mask is then collapsed into a single-pixel wide structure using medial axis skeletonization, and then extraneous branches are removed (in this case, any branches with a length less than 3 JCMT beam-widths).

\begin{figure}
    \vspace{-40pt}
	\includegraphics[width=\columnwidth]{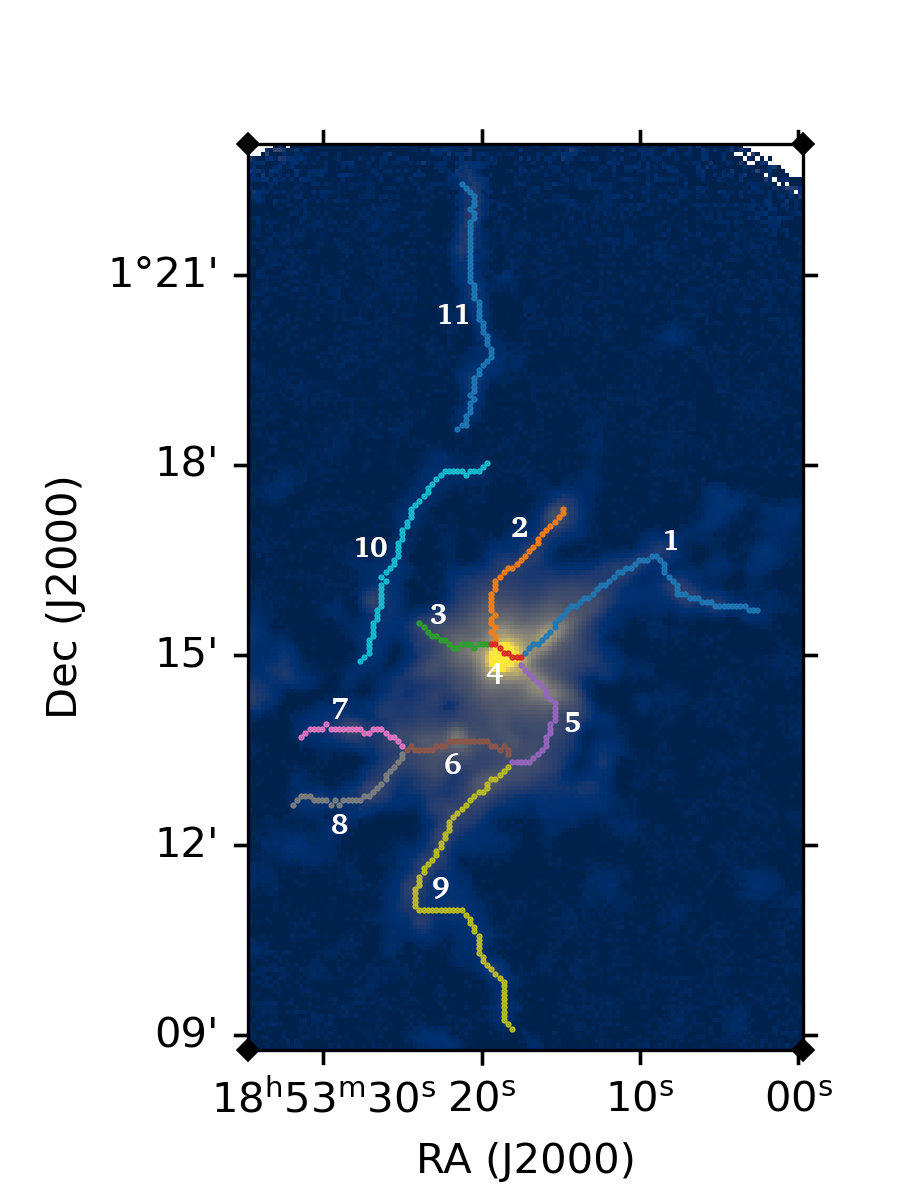}
    \vspace{-10pt}
    \caption{The filament skeleton generated by \textit{fil$\_$finder} from the 850$\mu$m Stokes I map of G34.26+0.15, with each filament represented as a series of pixels. Filaments are numbered from 1-11, as shown. Filaments 10 and 11 are considered as independent filaments with no physical connection to each other or the core skeleton.}
    \label{fig:fils}
\end{figure}

Note that we provided \textit{fil$\_$finder} with the entire Stokes I map, without the cuts described in Section \ref{sec:proc}, to allow the algorithm to create its own mask boundaries. As such, it produces some structure beyond our defined signal cut of ${I/\sigma_I} > 5$ for Stokes I. These were was removed, and the filaments manually separated and numbered, as shown in Figure \ref{fig:fils}.

As expected, this reveals several filaments extending out from the central core, which are particularly well-defined to the north and west of the central hub. Of note, however, are the branching filaments to the south: Filament 6 runs close to perpendicular to the central core, with its own sub-branches, Filaments 7 and 8. Notably, this coincides the local intensity peak noted in Section \ref{sec:bfields}, the significance of which is discussed in Section \ref{sec:feedback}.

Additionally, there are two filaments that are not part of the main skeleton structure. Although \textit{$fil\_finder$} considers these as two unconnected objects, it is likely they are tracing a single long filament, with a portion that has insufficient intensity to be distinguished from the background. Notably, this filament corresponds to the IRDC described by \citet{xu2016}, more commonly associated with the G34.43+0.24 star-forming region to the north of G34.26+0.15.

Filament 4 corresponds to the central hub of G34.26+0.15. Although this is not a filament per se, it is retained to maintain the full skeleton generated by \textit{$fil\_finder$}.

The G34.26+0.15 field was split, sorting each pixel based on its closest filament. In order to study the relation between the magnetic field orientations and local filament direction, the HRO and PRS techniques were performed again, but taking:

\begin{equation}
    \phi_{F} = \frac{\textbf{B} \times \textbf{F}}{\textbf{B} \cdot \textbf{F}}
	\label{eq:phi_fil}
\end{equation}

where $\textbf{F}$ is the tangent to the local filament. Equation \ref{eq:shape_factor} was again used to calculate what will be referred to as the filament alignment factor ($\xi_F$). This is analogous to $\xi_{I}$, with positive values corresponding to \textbf{B} and \textbf{F} being preferentially parallel, and negative values corresponding to \textbf{B} and \textbf{F} being preferentially perpendicular.

\begin{figure*}
\begin{minipage}[ht]{\textwidth}
    \centering
    \vspace{0pt}
    \begin{minipage}[t]{0.49\textwidth}
        \includegraphics[width=\textwidth]{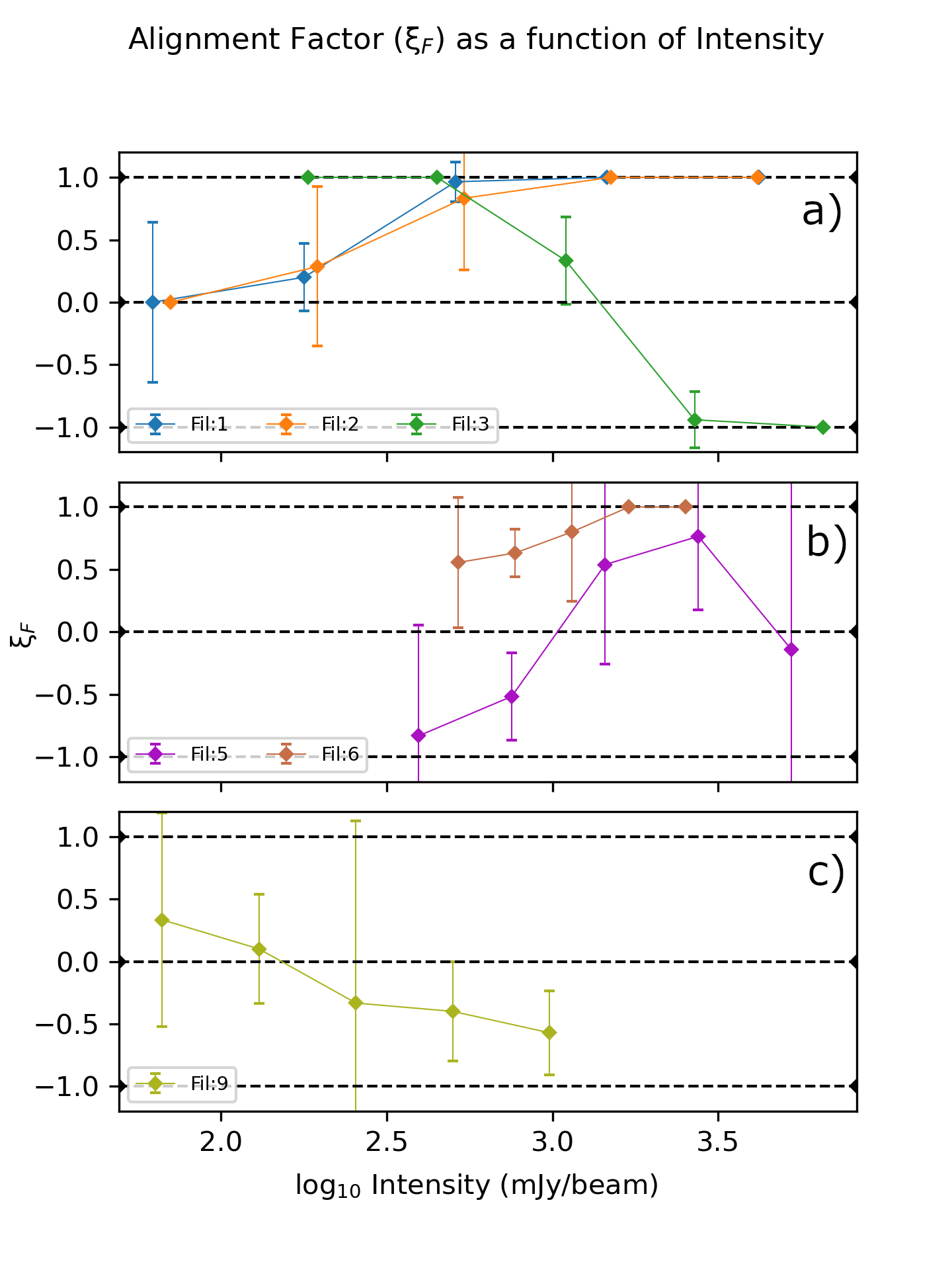}
        \vspace{0pt}
    \end{minipage}
    \begin{minipage}[t]{0.49\textwidth}
        \includegraphics[width=\textwidth]{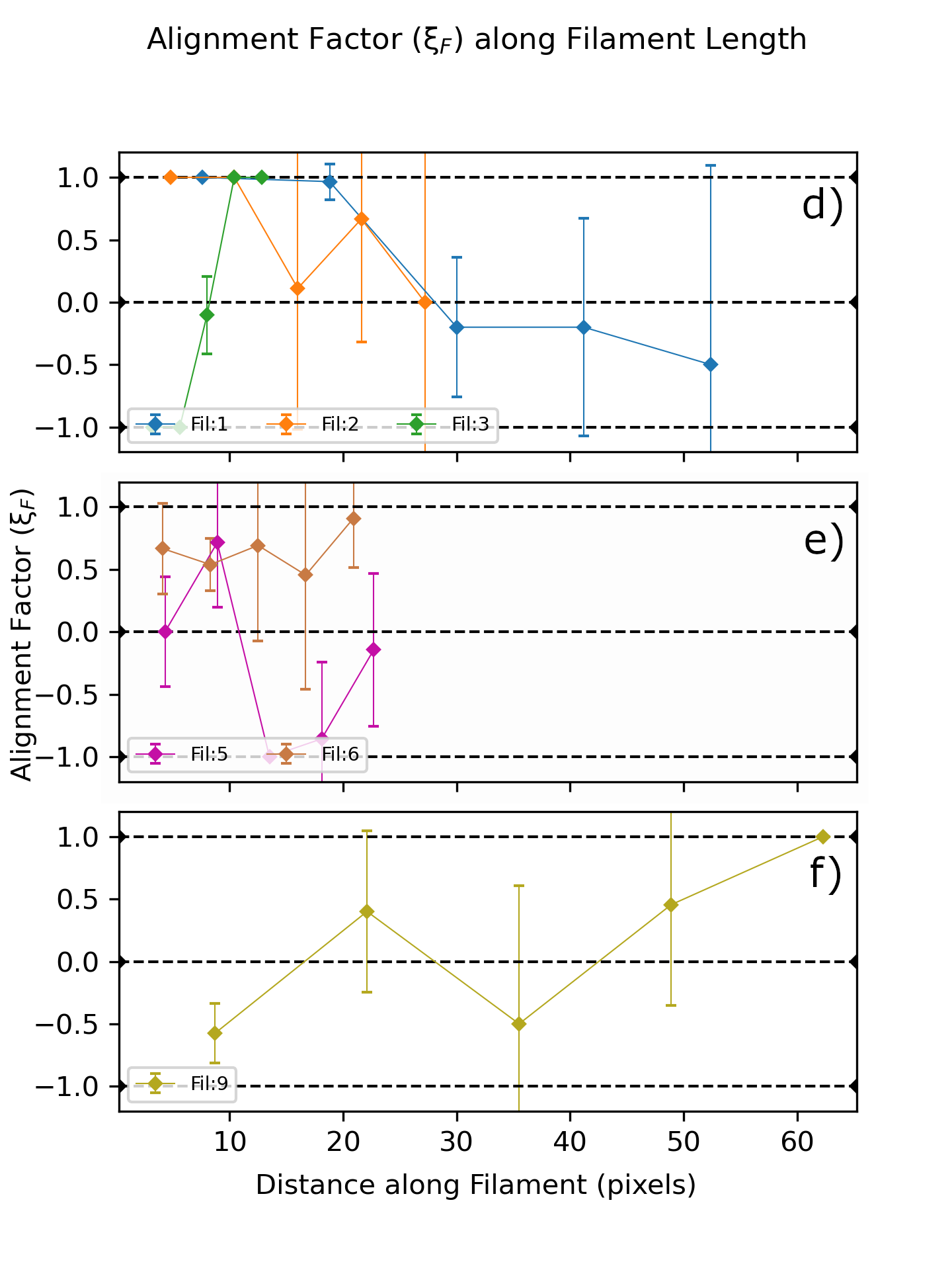}
        \vspace{0pt}
    \end{minipage}
    \caption{HROs showing the alignment factor $\xi_F$ for 850$\mu$m Stokes I intensity bins for various filaments across G34.26+0.15, as defined in Figure \ref{fig:fils} \textit{(Left: $\xi_F$ with respect to intensity, Right: $\xi_F$ with respect to position along the filament)}. The density gradients are defined as according to Equation \ref{eq:gauss_angle}. Errors are given as the combination of the standard errors of $A_e$ and $A_c$, as detailed in Equation \ref{eq:shape_err} The dotted lines show $\xi_F=1$ and $\xi_F=0$. The top row shows filaments associated with the central hub, the middle row shows filaments associated with the secondary hub, and the bottom row shows Filament 9, which is not directly associated with either.}
    \label{fig:HRO_fils}
\end{minipage}
\end{figure*}

The results of this are shown in Figure \ref{fig:HRO_fils} -- from this, the relationship between magnetic field orientation and density can be inferred for individual filaments. Six of the filaments have been omitted -- this is due to a lack of polarization angles for some filaments (Filaments 10 and 11) and the short length of others (Filaments 7 and 8). Filament 4 is also omitted, as previously noted it does not correspond to a physical filament. It should be noted that, as before, intensity is only a rough proxy for distance from the center of G34.26+0.15. For Filament 6 in particular, this proxy does not hold, with the greatest source of intensity being located at the midpoint of the filament.

The values of $\xi_F$ for Filaments 1 and 2 are notable for being very similar to the overall trend across G34.26+0.15, as seen in Figure \ref{fig:HRO_full}, increasing from an unaligned state to fully aligned-state at high intensities. Notably however, Filament 3 does not follow this trend, instead displaying the inverse: a strong parallel alignment at low intensities, which switches to a perpendicular alignment at high intensities, despite having a common origin with Filaments 1 and 2. This discrepancy is discussed in Section \ref{sec:feedback}. It also of note that Filament 2 has a clear peak in alignment when considering distance along the filament -- this corresponds to a local intensity peak.

Filament 6 is plotted here as a representative of the `wing' component of G34.26+0.15 (as defined in Section \ref{sec:velocity}). There is a strong parallel alignment here, with only a slight decrease at lower intensities. This is either due to a lack of information at lower intensities -- as the area encompassed by Filament 6 does not include any pixels below 300 mJy -- or due to the magnetic field instead being aligned to a different structure, as discussed in Section \ref{sec:feedback}.

Filaments 5 and 9 are also plotted, although both display a more inconsistent relationship between \textbf{B} and \textbf{F}. For Filament 5, this may be due to not being a `true' filament, instead being an artifact of \textit{$fil\_finder$} existing to connect the filamentary structures of the central and secondary hubs - this is plausible for a particular assumed morphology, as detailed in Section \ref{sec:kinematics}. This would explain the high parallel alignment close to the central hub, compared to the very sudden drop-off to an almost perpendicular alignment further along the filament. Filament 9 presents a somewhat different scenario, being by far the longest filament recorded. Notably, it coincides with the boundary of a H\textsc{ii} region detailed by \citet{xu2016}, the implications of which are discussed in Section \ref{sec:feedback}.

\begin{figure*}
\begin{minipage}[th]{\textwidth}
    \includegraphics[width=\textwidth]{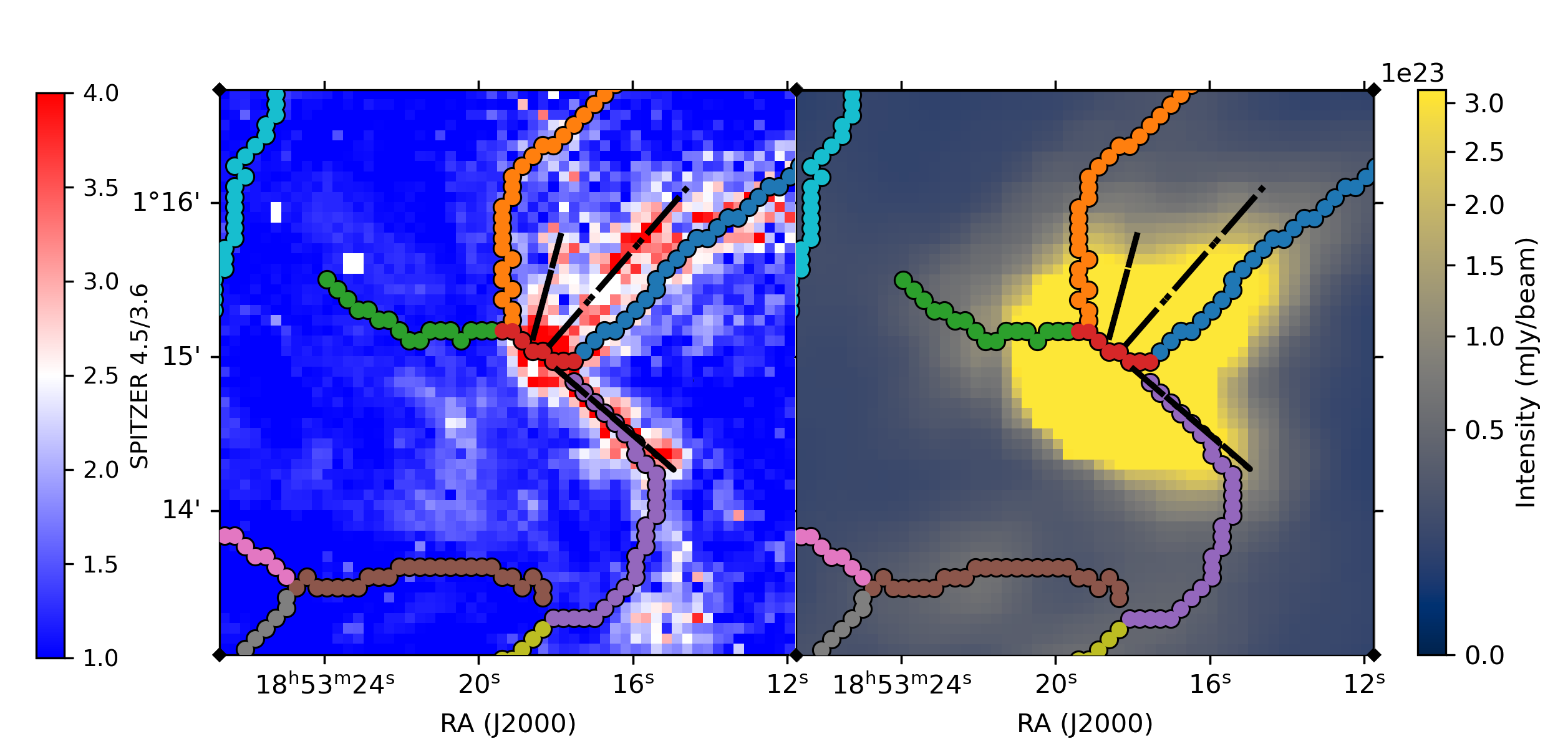}
    \vspace{-20pt}
    \caption{\textit{Left:} the filament structure detailed in Section \ref{sec:fils}, imposed on the flux ratio map of \textit{Spitzer} 4.5/3.6 \um\ emission. \textit{Right:} The same filament structure, imposed on the column density maps provided by the HOBYS survey \citep{motte2018}. In both cases, the black dashed lines indicate likely outflow directions.}
    \label{fig:fils_split}
\end{minipage}
\end{figure*}

\citet{liu2013} identify possible protostellar outflows across the plane-of-sky, originating from the hub of G34.26+0.15. In particular, they note strong \textit{Spitzer} 4.5\um\ emission surrounding the hub, indicative of shocked molecular H$_2$ \citep{takami2010}, as would be expected in the presence of an outflow. The outflows that \citet{liu2013} identify emanate from the north-western side of the central hub. In order to investigate whether there is any relationship between their proposed outflows, and our proposed filaments in this region, we repeated their analysis. We plot a \textit{Spitzer} 4.5/3.6 \um\ ratio map in Figure \ref{fig:fils_split}. This ratio is chosen to account for the effects of stellar contamination on the Spitzer 4.5\um\ band \citep{takami2010}.

As can be seen in Figure \ref{fig:fils_split}a, Filament 5 in particular seems to align very closely to a region of shocked material, which may explain its inconsistent $\xi_F$. Filaments 1 and 2, although present in an area of high 4.5/3.6 \um\ ratio, have spines offset from the loci of peak 4.5/3.6 \um\ ratio. Instead, they align much more closely with the dust (and so H$_2$) column density structures traced by Herschel HOBYS observations \citep{motte2018}, as shown in Figure \ref{fig:fils_split}b. While we cannot rule out the possibility that these filaments are outflowing, this correlation with column density structure, as well as the gravitational potential structure of the region (see Section \ref{sec:halves}), leads us to identify these filaments as likely infall structures onto the central hub.

\subsection{Velocity Analysis}
\label{sec:velocity}

\begin{figure*}
\begin{minipage}[ht]{\textwidth}
    \includegraphics[width=\textwidth]{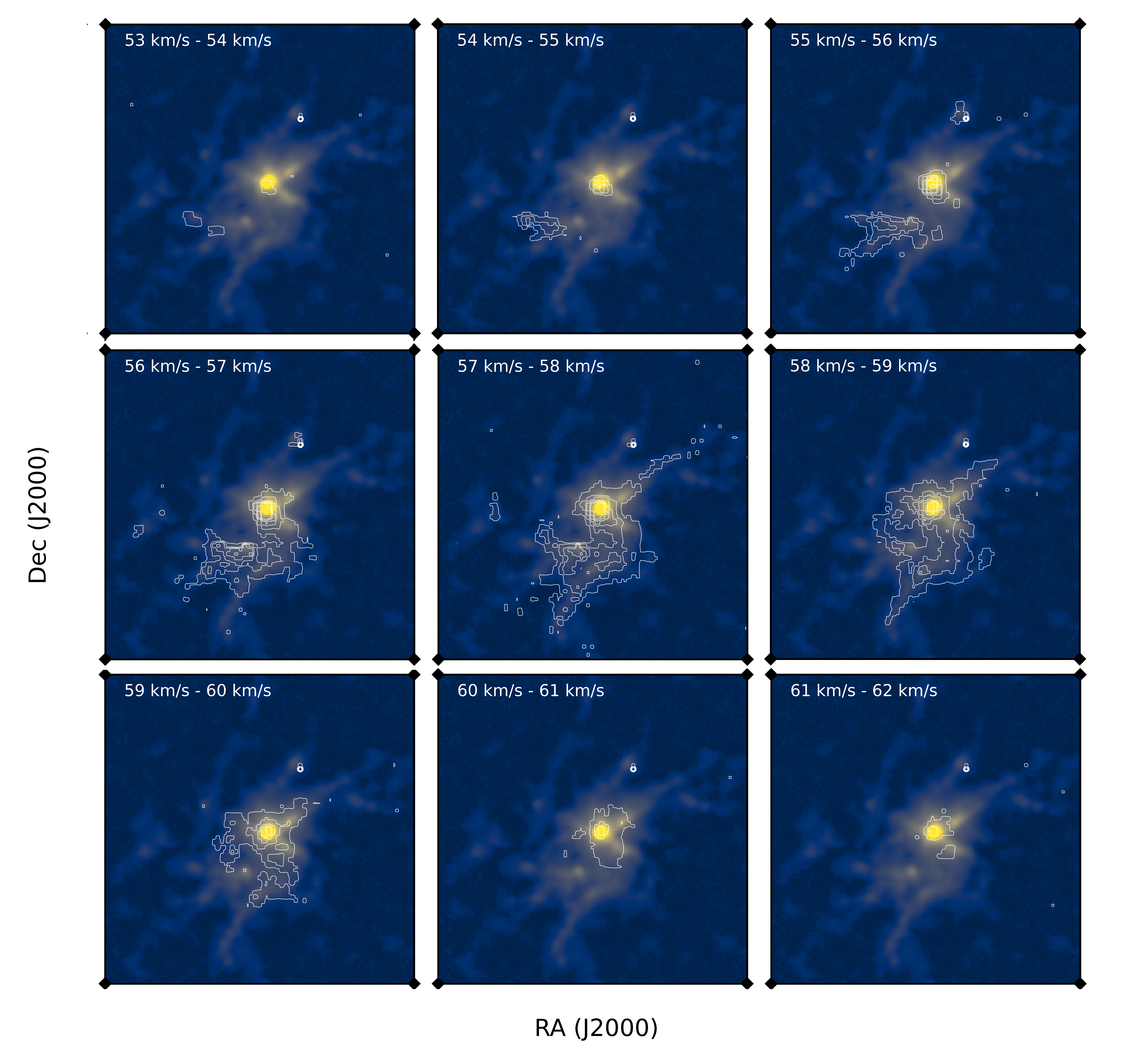}
    \vspace{-20pt}
    \caption{C$^{18}$O J=$3\to2$ channel map contours overlaid on the 850\um\ Stokes I map of G34.26+0.15. Channel widths are 1km/s. For the equivalent channel map for $^{13}$CO J=$3\to2$, see Appendix 1.}
    \label{fig:vels_all}
\end{minipage}
\end{figure*}

The hub-filament model suggests that gas within a molecular cloud is being channelled through filaments, flowing down towards the central hub (e.g. \citealt{pillai2020}). However, this particular scenario cannot be verified as occurring within G34.26+0.15 from magnetic field information alone (see Section \ref{sec:bfields}).

To supplement our magnetic field observations, we use observations of the C$^{18}$O J=$3\to2$ transitions, from the CO Heterodyne Inner Milky Way Plane Survey (CHIMPS, \citealt{rigby2016}) carried out with the Heterodyne Array Receiver Program (HARP) on the JCMT. Given that the frequency required to observe this transition is close to the 850\um\ used by POL-2 (329.331 GHz, or 910.31\um), it also has a very similar effective beam width of 14.2\arcsec. The data has a spectral velocity resolution of 0.5 kms$^{-1}$ after binning by the CHIMPS survey. Note that $^{13}$CO J=$3\to2$ data is also available -- this can be seen in Appendix \ref{sec:13co}, and was not used for analysis here.

These line data are summarized in Figure \ref{fig:vels_all}, binned to 1.0 kms$^{-1}$ between 53 kms$^{-1}$ and 62 kms$^{-1}$. It is notable that the majority of signal corresponds to a similar area as the POL-2 850\um\ data -- i.e. in the central 3 arc-minutes, and along the wing component of the region -- as this would suggest it is likely tracing the same material. The majority of G34.26+0.26 appears to have a characteristic velocity of 57-58 km/s, which agrees with previously determined values by \citet{xu2016}.

In order to more accurately determine the velocity characteristics, fitted the spectra with a Gaussian model, using the \textit{specutils} python package. This involves considering each pixel on the map as an independent spectrum (ignoring correlation within beam widths), and using an iterative process to constrain three parameters: mean velocity $v$, velocity dispersion $\sigma$ and peak intensity $A$. As the lines are relatively simple and isolated, we use the \textit{find$\_$lines$\_$threshold} and \textit{find$\_$lines$\_$derivative} functions within \textit{specutils} to estimate the mean line velocity, to use as the initial parameter in the fitting algorithm.

\begin{figure}
	\includegraphics[width=\columnwidth]{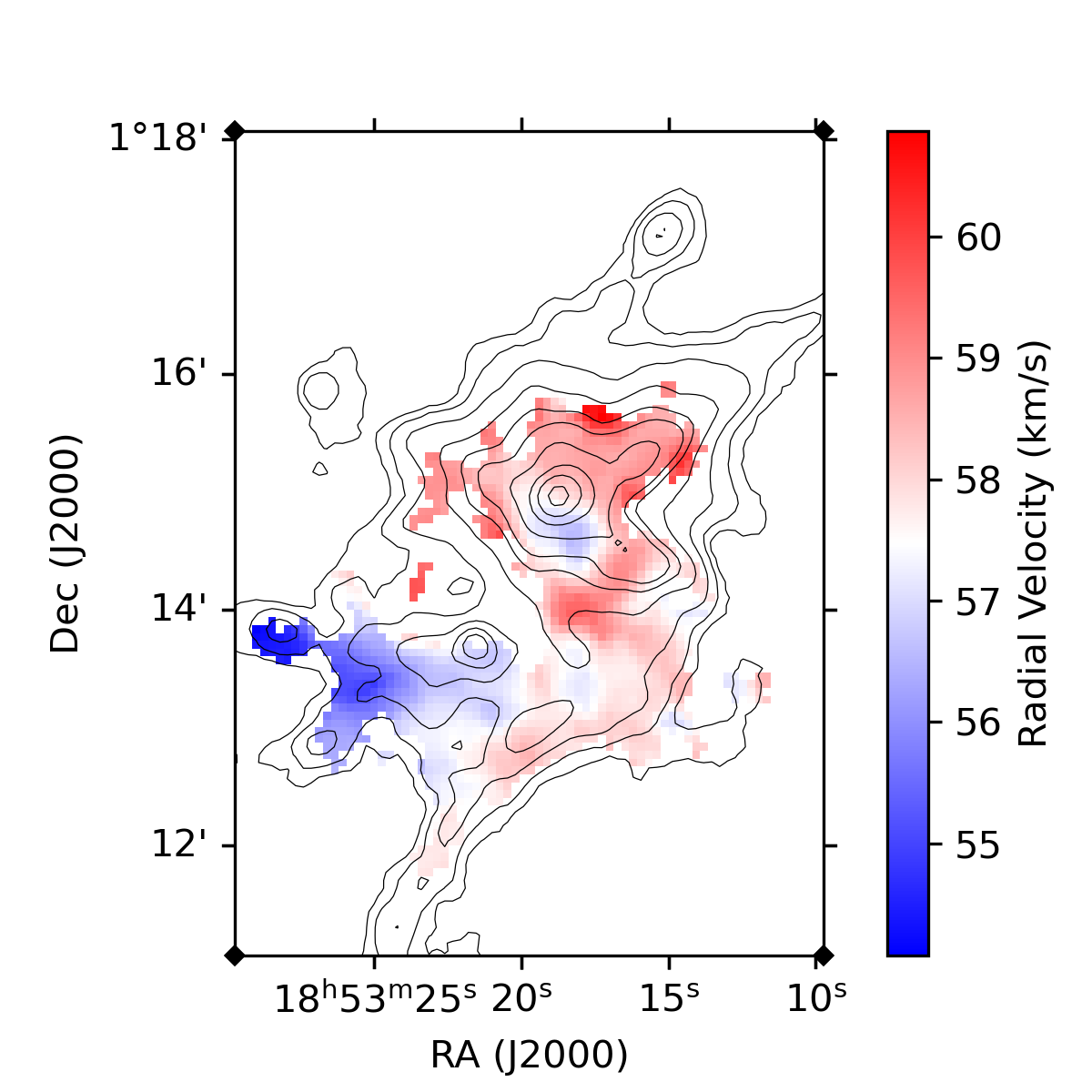}
    \caption{A mean velocity map of the central region of G34.26+0.15, derived from C$^{18}$O J=$3\to2$ HARP data. The contours shown are Stokes I emission at: 0.25\%, 0.5\%, 1\%, 2\%, 3.5\%, 5\%, 10\%, 20\%, 40\% and 80\% of the peak Stokes I intensity.}
    \label{fig:c18o_gauss}
\end{figure}

The mean velocity map produced by this method for C$^{18}$O J=$3\to2$ is shown in Figure \ref{fig:c18o_gauss}. Although Figure \ref{fig:vels_all} shows there is marginal detection of the extended filaments, here it is clear that the majority has an insufficient S/N to accurately fit a Gaussian curve to. Visual inspection of individual spectra confirms this. However, the central hub and the wing section are well-detected, and a clear velocity gradient across the region can be seen. This ranges from +54 km/s at the very eastern edge of the wing, to +61 km/s to the north of the central hub (although the majority of data is between +55 and +59 km/s), with a systematic velocity of +57--58 km/s.

The smooth velocity change across the region suggests that everything observed is physically associated, rather than material at significantly different distances. We see an increasing velocity gradient towards the central hub, as would be expected if material is flowing towards (or possibly away from) the centre. A similar effect has been observed in other hub-filament systems, such as Monoceros R2 \citep{trevinomorales2019}. This behaviour is particularly prominent in Filament 1, as seen in Figure \ref{fig:vels_all}.

The possible second hub discussed in Section \ref{sec:fils} complicates this structure. Although the secondary hub is sufficiently gravitationally dominated to produce a massive star \citep{hunter1998}, the C$^{18}$O velocity information instead shows a velocity gradient through this hub from east to west. This seems to suggest a flow of gas through the system, with material increasing in velocity towards the center of the field.

An alternative explanation for the observed velocity gradient involves the line-of-sight structure of the filament system. See Section \ref{sec:feedback} for further details.

\subsection{Alfven/DCF Analysis}
\label{sec:dcf}

A major limitation of using dust emission to measure magnetic field directions is the lack of associated field strength measurements. However, the plane-of-sky field strength can be estimated using the Davis-Chandrasekhar-Fermi (DCF) method (\citealt{davis1951}, \citealt{chandra1953}). This method assumes that small-scale random variations in magnetic field orientation is caused by Alfvén waves formed by turbulence by perturbing the field, and that the relationship is sub-Alfvénic (i.e. the magnetic field is able to react to changes in the gas medium). The stronger the magnetic field, the lower variation should be as turbulent motion is less able to perturb the magnetic field.

Traditional methods of DCF have been noted to potentially overestimate the magnetic field strength due to not considering multiple turbulent cells through the LOS, instead integrating through these components. \citet{choyoo2016} note that this overestimation can be compensated for by using a modified version of the DCF equation:

\begin{equation}
    B_{POS} = Q \sqrt{4\pi\rho } \frac{\delta V_c}{\delta\theta},
	\label{eq:dcf}
\end{equation}

where $\delta V_c$ is the dispersion of centroid velocities across the given region, and $\rho$ is the gas density in the region. \textit{Q} is a constant of order unity, that \citet{choyoo2016} determine to be between $\sim$0.7 and $\sim$1.0 through numerical simulations -- we take $Q$ to be 0.7 in our analysis. We combine this form of the DCF with the unit conversions detailed by \citet{crutcher2004}:

\begin{equation}
    B_{POS} \approx 44 Q \sqrt{n({\rm H}_2) (\rm cm^{-3})} \frac{\delta V_c (kms^{-1})}{\delta\theta (^\circ)} \mu G,
	\label{eq:dcf_alt}
\end{equation}

where $n({\rm H}_2)$ is gas volume density and $d\theta$ is the magnetic field angle dispersion.



We take into account of the effect of large-scale structure in the magnetic field
 on  $\delta\nu$, using the `structure function' $\Delta \Phi^2 (\ell)$ method detailed by \citet{hildebrand2009}. The structure function can be calculated by considering the average angle difference between all magnetic field vectors at a distance $\ell$ from one another. This can be approximated as having three components::

\begin{equation}
    \Delta \Phi^2 (\ell) = b^2 + m^2\ell^2 + \sigma^2_M(\ell)
	\label{eq:struc_func}
\end{equation}

where $\sigma^2_M(\ell)$ is the contribution from uncertainties on the magnetic field vectors and $b$ is the turbulent contribution. $m^2$ is the fitted slope factor on $\ell^2$. \citet{hildebrand2009} show that:

\begin{equation}
    {\delta\theta} = \frac{b}{\sqrt{2-b^2}} \simeq \frac{b}{\sqrt{2}}
	\label{eq:b}
\end{equation}

\begin{figure*}
\begin{minipage}[ht]{\textwidth}
    \includegraphics[width=\textwidth]{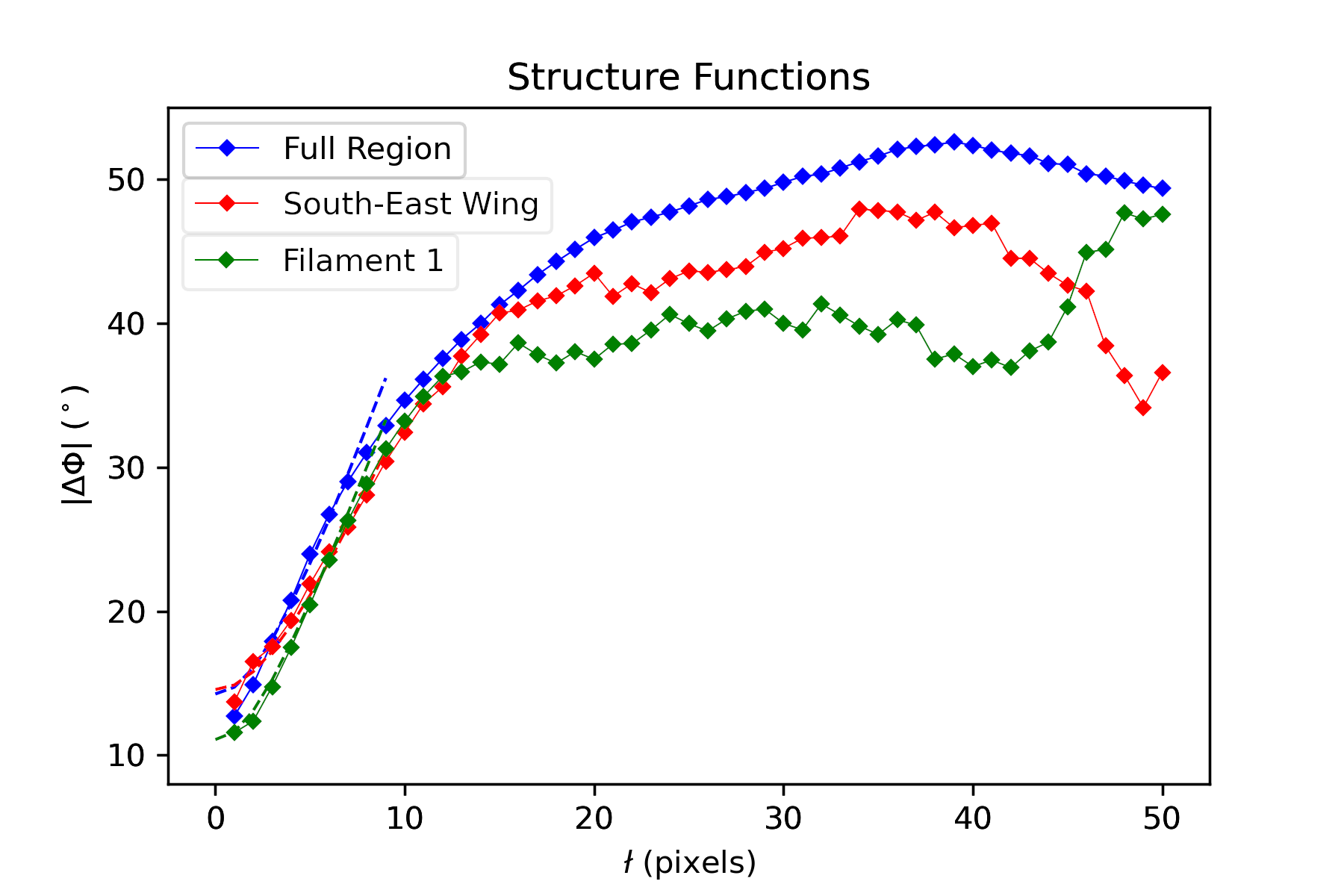}
    \vspace{-20pt}
    \caption{Structure functions of the magnetic field for different components of G34.26+0.15. Filament 1 is defined by the area used for this filament in Section \ref{sec:fils}, while the South-East Wing is defined by the velocity map shown in Figure \ref{fig:c18o_gauss}.}
    \label{fig:struc_func}
\end{minipage}
\end{figure*}

In practice, $b^2$ can be determined by plotting the structure function (corrected for the contribution of $\sigma^2_M(\ell)$ by subtracting it for all values of $\ell^2$) and fitting to determine $m^2$. Figure \ref{fig:struc_func} shows the structure function for the G34.26+0.15 field. Three curves are shown: one for the full field, one for the wing component largely corresponding to Filament 6 in the south-east half, and one for Filament 1 in the north-west half. These were isolated as regions with sufficient reliable velocity data to calculate $\delta V_c$. In each case, the corresponding dashed lines represents a linear fit between $\ell^2$ and $\Delta \Phi^2 (\ell)$ using the Equation \ref{eq:struc_func}. Only values of $\ell$ below 9 pixels are used to reduce the effect of contamination by large-scale structures.

From the POL-2 data, we thus calculate $b$, which can be used to determine ${\delta\theta}$ using Equation \ref{eq:b}. With the velocity data from HARP detailed in Section \ref{sec:velocity} , we calculate $\delta V_c$ as described previously using the method by \citet{choyoo2016}. Seperately, we are able to able to estimate the Alfven Mach Number ($M_A = \sqrt{3} \frac{\delta\theta}{Q}$, \citealt{hwang2021}) -- this is a measure of the relative strength of magnetic field compared to turbulent effects. The Alfven Mach Number was calculated for each region, the results of which are summarized in Table \ref{tab:dcf}.

\begin{table}
 \caption{DCF Statistics across G34.26+0.15}
 \label{tab:dcf}
 \begin{tabular}{lccccc}
  \hline
  Region & $b$ & $\delta V_c$ & $M_A$ & n(H$_2$) & $B_{POS}$\\
   & ($^\circ$) & (kms$^{-1}$) &  & (cm$^{-3}$) & (mG) \\
  \hline
  G34.26+0.15 (Full) & 14.3 & 3.4 & 0.61 & 26800 & 0.53\\
  N-W Filament 1 & 11.0 & 5.2 & 0.47 & 40500 & 1.12\\
  S-E Wing & 14.6 & 2.7 & 0.62 & 148000 & 1.39\\
  \hline
 \end{tabular}
\end{table}

For a full DCF analysis, we require the volume density of the gas $n({\rm H}_2)$ within the region, which is typically estimated from observed column density. We use column density maps created by the HOBYS survey \citep{nguyen2011}, which are created from observations from the SPIRE and PACS instruments on-board the \textit{Herschel} Space Observatory through SED fitting: This is suitable for the wing region and Filament 1, but there is a fitting issue in the central region -- here, it is filled with a constant value. This is  due to oversaturation of the SPIRE instrument at this position. 


In order to convert these column densities into a volume density, there are  certain assumptions to be made. We treat the full region as a ellipsoidal prism, with the LOS depth taken to be the average of the POS height and width. Filament 1 is treated as thin cylinders orientated along the POS, with its depth assumed to be equal to their width, and for comparison the wing region is assumed to have same depth. Using the average column density across each region, we estimate the volume density, and thus estimate $B_{POS}$ using Equation \ref{eq:dcf_alt}. The resulting values are presented in Table \ref{tab:dcf}.

Although the values presented for $B_{POS}$ for both the Filament 1 and the wing component are distinctly higher than that of the full region, we note that DCF is only valid as an order of magnitude estimate of the magnetic field strength. This is particularly true in the case of G34.26+0.15, as we lack information about the 3D density structure of the molecular cloud, and thus our assumption about volume densities may be inaccurate. Similarly, although the value of $B_{POS}$ for the south-east wing component appears higher than that of the Filament 1, in reality this is likely to be an upper limit, as true volume density may be much lower.

\section{Discussion}
\label{sec:discussion}

\subsection{G34.26+0.15 Halves Magnetic Field Comparison}
\label{sec:halves}

As discussed briefly in Section \ref{sec:bfields} and shown in Figure \ref{fig:g34_map}, there appears to be two distinct `halves' to the extended G34.26+0.15 region -- the north-west half which displays apparently radial magnetic field lines around the central hub, and the south-east half which displays a more complex structure as detailed in Section \ref{sec:bfields}. In this section, we will quantify this difference, making reference to the filamentary structure seen in Section \ref{sec:fils}.

\begin{figure}
	\includegraphics[width=\columnwidth]{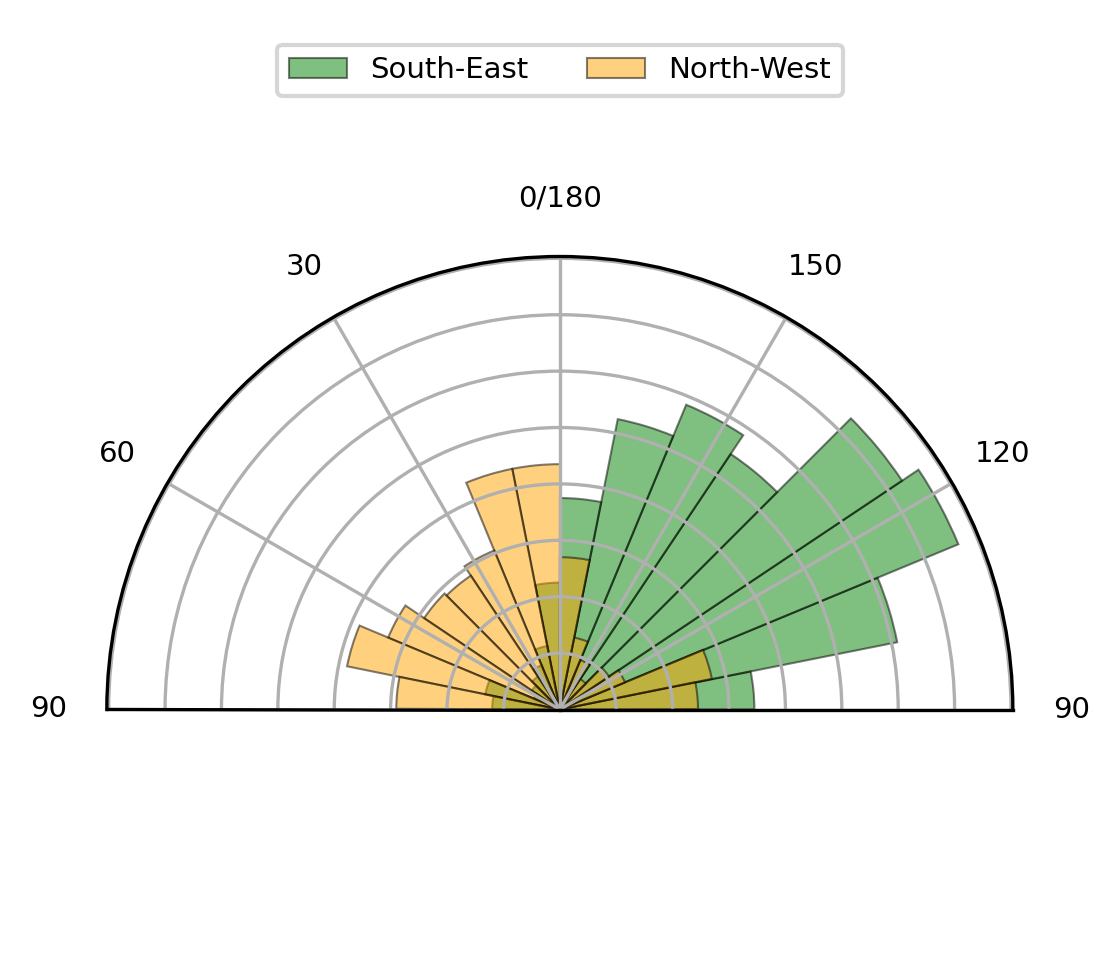}
    \vspace{-60pt}
    \caption{A radial plot of the magnetic field vectors across G34.26+0.15, as measured counter-clockwise from north. Note that angles are constrained from 0$^\circ$ and 180$^\circ$, taking into account the 180$^\circ$ ambiguity in polarization angle (see Section \ref{sec:datared}).}
    \label{fig:polar_plot}
\end{figure}

Figure \ref{fig:polar_plot} shows a polar plot, with magnetic field directions for both halves (as defined in Figure \ref{fig:g34_map}) overlaid. It is clear that the halves have a very different preferred orientation, with the north-west half having an average angle of 104$^\circ$ and the south-east half having an average angle of 58$^\circ$ east of north. The north-west region also has a somewhat higher dispersion around the average direction, with a standard deviation of 51$^\circ$, compared to 40$^\circ$ for the south-east region.

This difference here again suggests that the south-east does not exhibit the same radial filamentary structure as seen in the northern half. As discussed in Section \ref{sec:fils}, the south-east half contains a distinctly separate filamentary structure, encompassing Filaments 6-8, which has a magnetic field structure that aligns parallel with Filament 6. However, the average magnetic field angle around Filament 6 is closer to 90$^{\circ}$, which is significantly different to average of the 58$^\circ$ for the south-east half.

The majority of the magnetic field vector contribution in the south-east half arises from a large area approximately an arc-minute in width, in which is there is  markedly less 850\um\ emission than the surrounding structure. This decrease in 850\um\ emission coincides with a region of strong 8\um\ and 6cm (C-band) emission, as shown in Figure \ref{fig:spitzer} using \textit{Spitzer} and VLA data \citep{crossley2007}. As discussed below, this is most likely due to the presence of the G34.26+0.15 H\textsc{ii} Region D (\citealt{reid1985}, \citealt{gaume1994}). As noted by \citet{liu2013}, the \textit{Spitzer} 8\um\ PAH emission forms a distinct `arc' around the 850\um\ void; the C-band emission shows similar structure, suggesting this is the boundary of the H\textsc{ii} region.

As mentioned in Section \ref{sec:dcf}, we see relatively constant magnetic field strengths of order $\sim$1 mG across both halves of G34.26+0.15 , if we take Filament 1 as representative of the north-west half and the wing component as representative of the south-east half. However, we do see somewhat different values of M$_{A}$ in the two halves, with 0.47 for the south-east and 0.62 for the north-west. This is a measure of the relative dynamic importance of the magnetic fields compares to turbulent gas motions. Given our use of the structure function method \citep{hildebrand2009}, this should be free of large-scale dispersion effects, and thus this suggests that the north-west half has a magnetic field that is weaker compared to turbulent effects than the south-east half. However, this may also be due to the north-west half being gravitationally dominated, with both magnetic and turbulent effects being dynamically much less important.

\begin{figure}
	\includegraphics[width=\columnwidth]{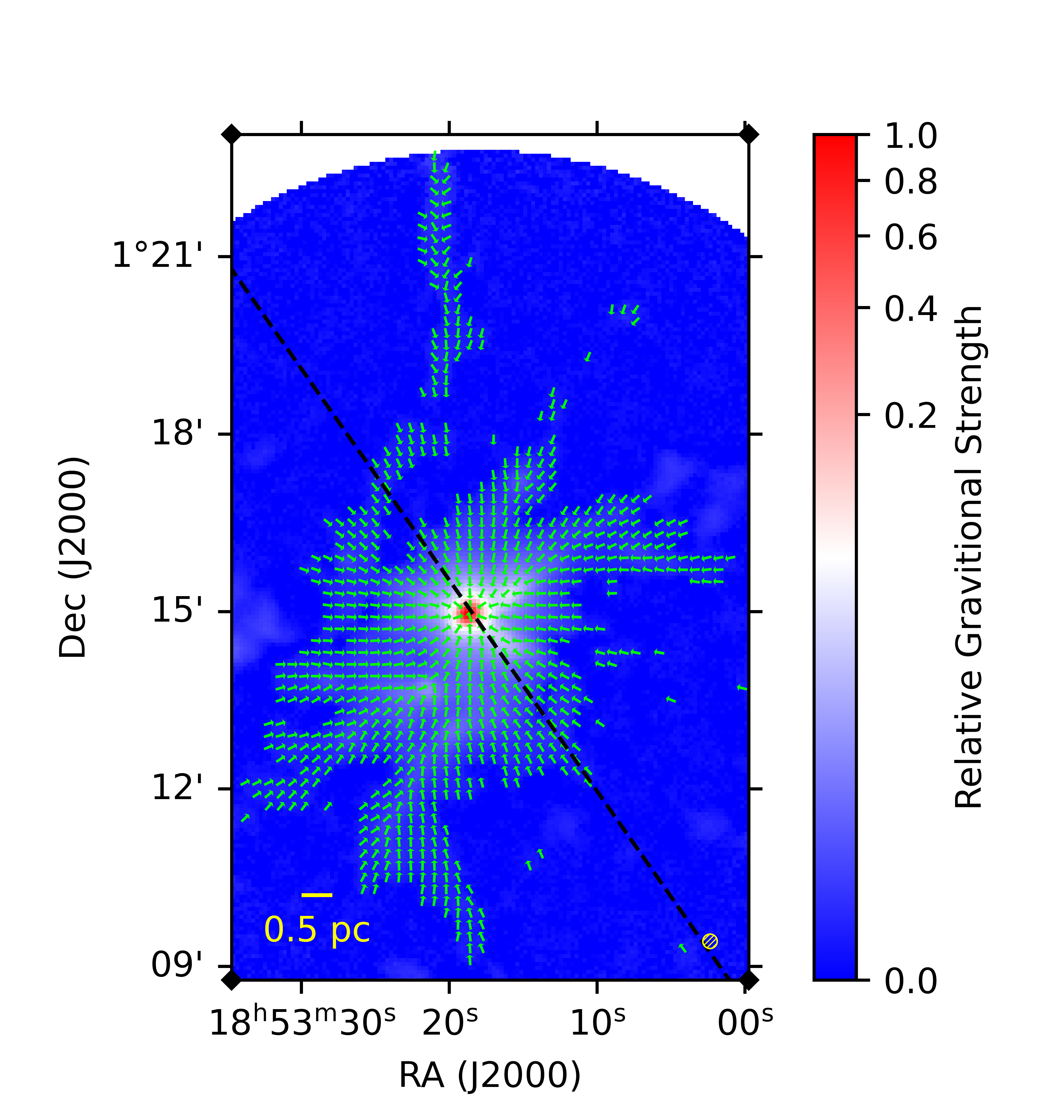}
    \caption{A map of relative gravitational field strength across G34.26+0.15, derived using the Stokes I intensity as a proxy for density. The gravity vectors are shown in green. The dotted black line represents the split between north-west and south-east used for the analysis in Figures \ref{fig:polar_plot} and \ref{fig:HRO_split}.}
    \label{fig:grav}
\end{figure}

To verify this, we independently analysed the two halves of G34.26+0.15, using the same HRO technique as detailed in Section \ref{sec:HROs}. Additionally, we considered the gravitational field vectors in G34.26+0.15 -- for this, we used intensity as a proxy for density. As described by \citet{wang2020}, the gravitational field vector for each pixel \textit{i} can be calculated by summing the gravitational force contribution from all pixels \textit{j} in the field in the form:

\begin{equation}
    \textbf{G}_{i} = I_i \sum^{n}_{j=1} \frac{I_j}{r^2_{i,j}} \hat{r},
	\label{eq:grav_vector}
\end{equation}

where \textit{I$_i$} and \textit{I$_j$} are the 850\um\ intensities of pixels \textit{i} and \textit{j}, \textit{n} is the total number of pixels in the field, \textit{r$_{i,j}$} is the distance between pixels \textit{i} and \textit{j}, and \textit{$\hat{r}$} is the associated unit vector. The resulting vectors are shown in Figure \ref{fig:grav} -- this method only gives the relative gravitational strength, rather than the absolute field strength. It is clear from these vectors that the central hub dominates the gravitational structure of G34.26+0.15, with most gravity vectors pointing towards the hub with only minor deviations for local structures. This is similar to the arrangement seen in NGC 2264C by \citet{wang2024}, which is a similarly gravity-dominated high-mass star-forming region.

The alignment of the magnetic field to the local gravity vectors was considered using a gravitational shape factor ($\xi_G$), calculated using Equation \ref{eq:shape_factor} but taking:

\begin{equation}
    \phi_{G} = \frac{\textbf{B} \times \nabla\textbf{G}}{\textbf{B} \cdot \nabla\textbf{G}},
	\label{eq:phi_grav}
\end{equation}

where $\nabla \textbf{G}$ is the direction of the local iso-gravity contour, perpendicular to \textbf{G}.

\begin{figure*}
\begin{minipage}[th]{\textwidth}
    \centering
    \vspace{0pt}
    \begin{minipage}[t]{0.49\textwidth}
        \includegraphics[width=\textwidth]{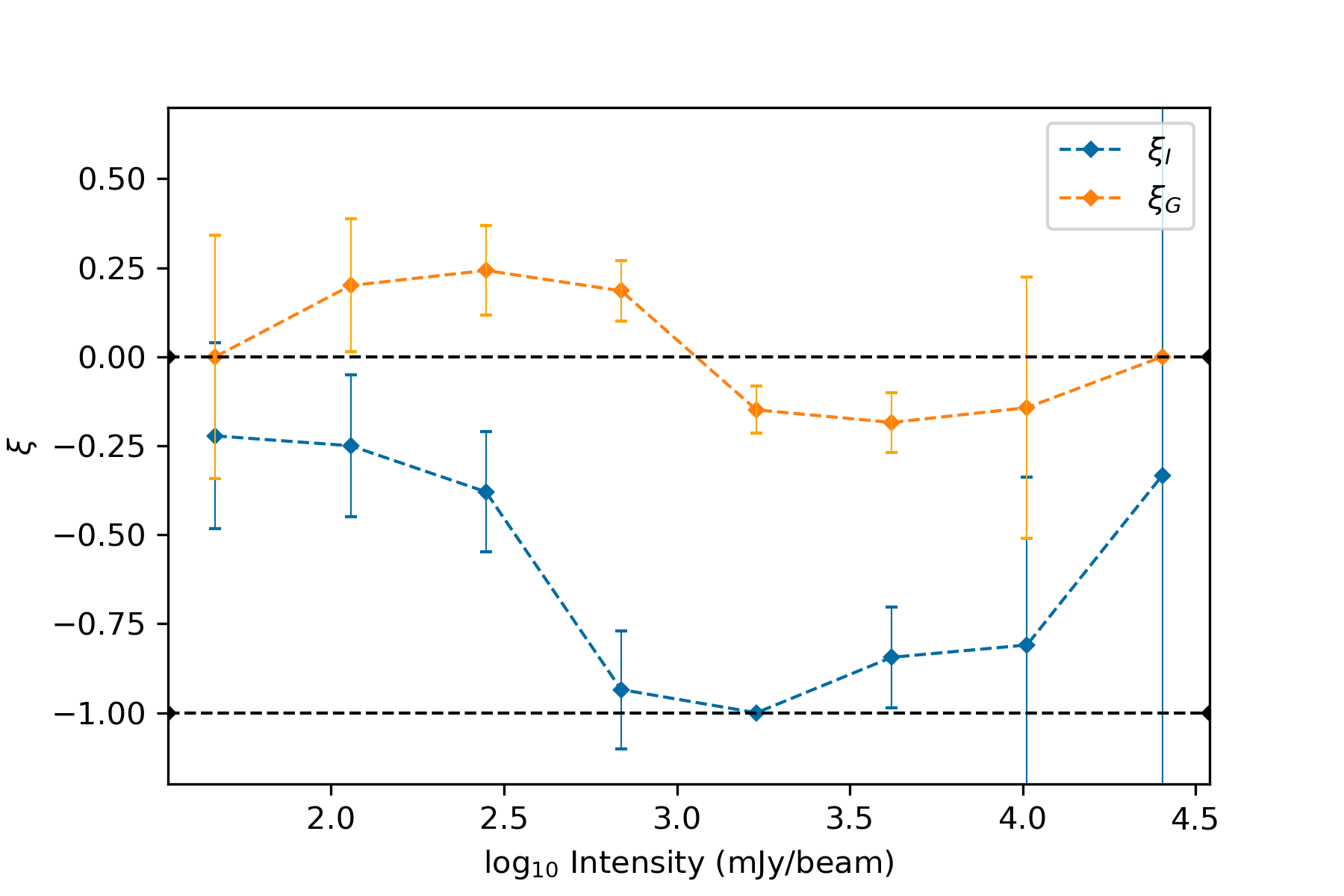}
        \vspace{0pt}
    \end{minipage}
    \begin{minipage}[t]{0.49\textwidth}
        \includegraphics[width=\textwidth]{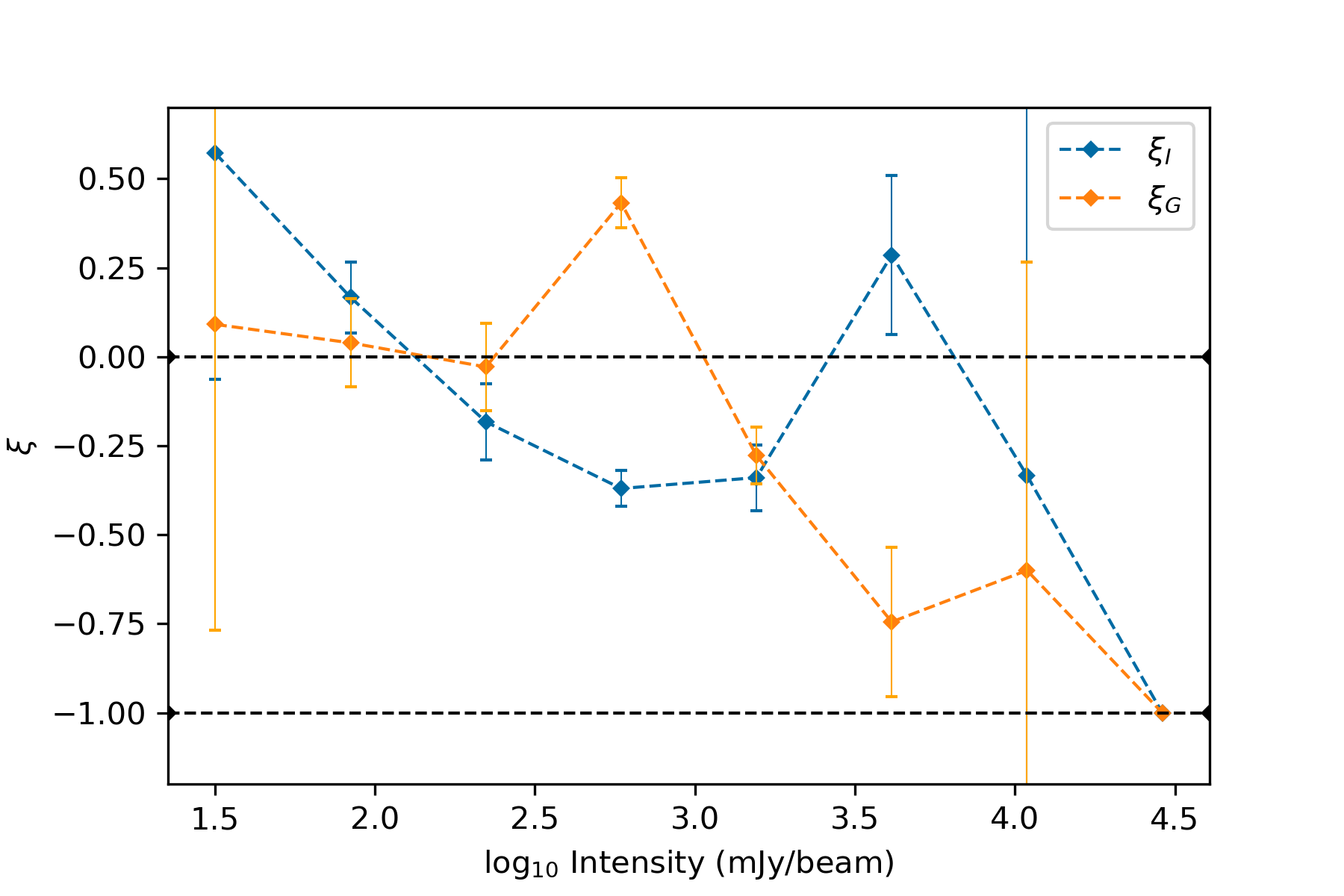}
        \vspace{0pt}
    \end{minipage}
    \caption{HROs showing $\xi_I$ and $\xi_G$ for 850$\mu$m Stokes I intensity bins for the two halves of the G34.26+0.15 field (\textit{Left: north-west, Right: south-east}, as defined in Figure \ref{fig:spitzer}). The density gradients are defined as according to Equation \ref{eq:gauss_angle}, and gravity vectors are taken to be the local direction of gravity according to \ref{eq:grav_vector}. Errors are given as the combination of the standard errors of $A_e$ and $A_c$, as detailed in Equation \ref{eq:shape_err}. The dotted lines show $\xi=1$ and $\xi=0$.}
    \label{fig:HRO_split}
\end{minipage}
\end{figure*}

The results of this are shown in Figure \ref{fig:HRO_split}, for both the north-west and south-east halves, the separation of which is also shown in Figures \ref{fig:grav} and \ref{fig:g34_map}. The distribution of $\xi_{I}$ is distinctly different for each half: the magnetic field in the north-west half is largely aligned along the intensity gradient with an entirely negative $\xi_I$ at all intensities, while the magnetic field in the south-east is differently aligned at different intensities, suggesting a more complex structure. We interpret this as being due to the presence of H\textsc{ii} region D, with this dominating the magnetic field structure in the south-east, while the north-west is gravity-dominated, with strong alignment of magnetic fields along intensity gradients in the same manner as seen using $\xi_F$ for Filaments 1 and 2.

The finding that the north-west half is gravity-dominated while the SE is supported by the calculated values of $\xi_G$. $\xi_G$ in the north-west largely mimics $\xi_I$, rising and falling at the same intensities. However, in general, $\xi_G$ is more positive than $\xi_I$ -- this is likely due to density structures and gravity direction misaligning at low densities, and the arrangement of the magnetic field around Filament 5 (see Section \ref{sec:kinematics}). In the south-east, however, $\xi_I$ and $\xi_G$ are similar at high and low intensities, but extremely different between 10$^{2.5}$ mJy and 10$^{3.5}$ mJy. This corresponds to the 850\um\ intensities which H\textsc{ii} region D spans, suggesting the region has significantly affected the magnetic field structure.

It may be possible that that the difference in both $\xi_G$ and $\xi_I$ between the two halves is due to differing magnetic contribution. This is supported by the similar difference in $\xi_G$ reported by \citet{wang2024} between NGC 2264C and NGC 2264D, which they partially attribute to a greater magnetic field contribution in NGC 2264D. However, they do not observe the same distinct difference in $\xi_I$ between NGC 2264C and NGC 2264D which we see in the two halves of G34.26+0.15, instead seeing an arrangement similar to G34.26+0.15's north-west half for both NGC 2264C and -D, with the magnetic field increasing aligning along density structures at higher intensities. Thus, we can again attribute our difference in $\xi_G$ and $\xi_I$ to the presence of H\textsc{ii} region D, which lacks a counterpart in NGC 2264.

\subsection{H\textsc{ii} Region D Feedback}
\label{sec:feedback}

\begin{figure}
	\includegraphics[width=\columnwidth]{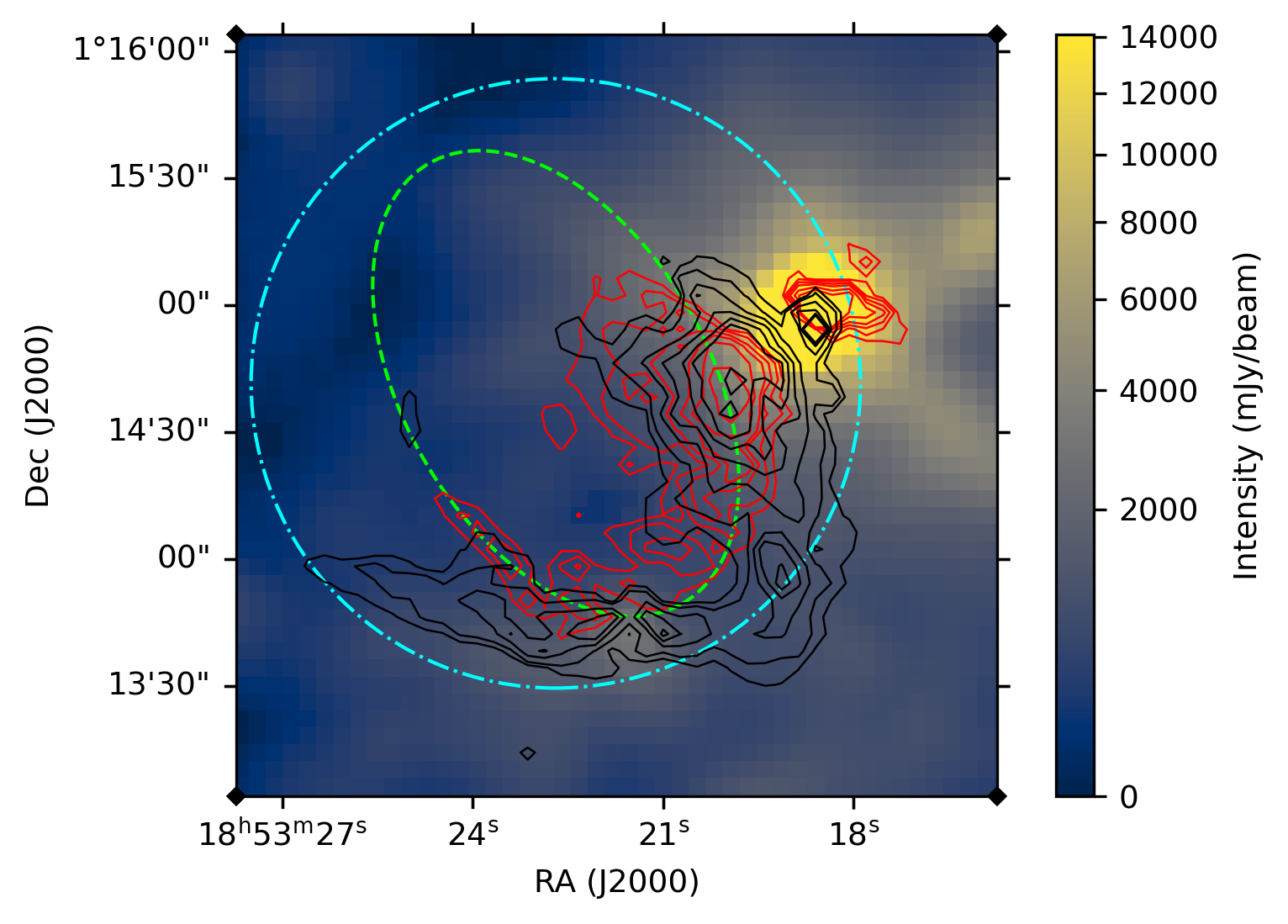}
    \caption{850\um\ Stokes I emission of central region of G34.26+0.15. Red contours show C band emission as observed by the VLA (\citealt{crossley2007}, contours levels of 1\%, 2\%, 3.5\%, 5\%, 7.5\% and 10\% of peak emission) in which H\textsc{ii} Region D is visible, and black contours show the \textit{Spitzer} 8\um\ emission, corresponding to the PAH emission which is proposed to be associated with the PDR of Region D (Contour levels of 7.5\%, 10\%, 15\%, 20\%, 25\% and 30\%). The cyan circle represents the bubble from the \citet{simpson2012} catalog, and the green ellipse represents our proposed shape of HII Region D, which is used in Figure \ref{fig:circle_spit}.}
    \label{fig:spitzer}
\end{figure}

The extended synchrotron emission shown in Figure \ref{fig:spitzer} corresponds with both the \textit{Spitzer} 8\um\ data, and with one of the `bubbles' identified by \citet{simpson2012}. The Simpson catalogue records infrared `bubbles' -- circular or partial arc-shaped features -- as identified by eye. Although the catalogue does not identify the nature of these features, it suggests that many are caused by shock front heating and expanding H\textsc{ii} regions \citep{simpson2012}. In the case of H\textsc{ii} Region D, we note that both the associated Simpson bubble and the C-band VLA emissions are significantly more extended than the 21cm VLA emission used to define the region by \citet{reid1985}. It is likely this is due to the original field-of-view of the \citet{reid1985} observations being insufficiently wide to capture the full extent of Region D.

Additionally, there is some ambiguity as to whether Region D is a single H\textsc{ii} region or multiple regions within close proximity. Although we interpret the structure as a single arc shape, it is plausible that is it actually composed of two or more H\textsc{ii} regions, with multiple driving sources. This ambiguity is due to the lack of identified driving stars within the vicinity: the closest large source identified is a B-type star. However, this star is at the very south edge of the H\textsc{ii} region(s), and has previously been instead associated with the UCH\textsc{ii} region G34.24+0.13 \citep{hunter1998}. We will assume Region D is a single H\textsc{ii} region for the remainder of this discussion, in the absence of evidence to the contrary.

Across H\textsc{ii} Region D, the magnetic field is close to linear, with a preferred orientation of $\simeq$35$^{\circ}$. The notable exception to this orientation is the afore-mentioned filamentary structure around the secondary hub of G34.26+0.15, although the field here is still orientated along the north-east/south-west. The presence of this H\textsc{ii} region would also explain the magnetic field alignment seen in Filament 3 (see Figure \ref{fig:spitzer}). As noted in Section \ref{sec:fils}, in the north-west half where we see a radial magnetic field, the field aligns parallel to Filaments 1 and 2  at higher densities. Along Filament 3, however, the magnetic field instead becomes aligned perpendicular to the filament at high densities. This filament traces the northern boundary of the \textit{Spitzer} 8\um\ emission -- thus it is likely that Filament 3 is not a self-gravitating filament as would be expected in a hub-filament system, but an overdensity caused by the compression of gas by the expanding H\textsc{ii} region.

Of particular interest is the mechanism by which the magnetic field has become aligned along the boundary of the H\textsc{ii} region, rather than the density and gravitational structures. The expansion of H\textsc{ii} Region D has likely compressed the surrounding gas, as suggested by the emission shown in Figure \ref{fig:spitzer}. Here, it can be seen that the 8\um\ emission is offset radially from the the C-band emission. 8\um\ emission is typically associated with hot dust (as opposed to the cold dust traced by 850\um\ emission) and C-band emission is associated with synchrotron emissions. Given that the 8\um\ emission is particularly bright in this area -- an order of magnitude higher than the rest of the G34.26+0.15 field -- we suggest that this emission is result of the expansion of the H\textsc{ii} region compressing the gas around it, forming a shock front.

\begin{figure}
	\includegraphics[width=\columnwidth]{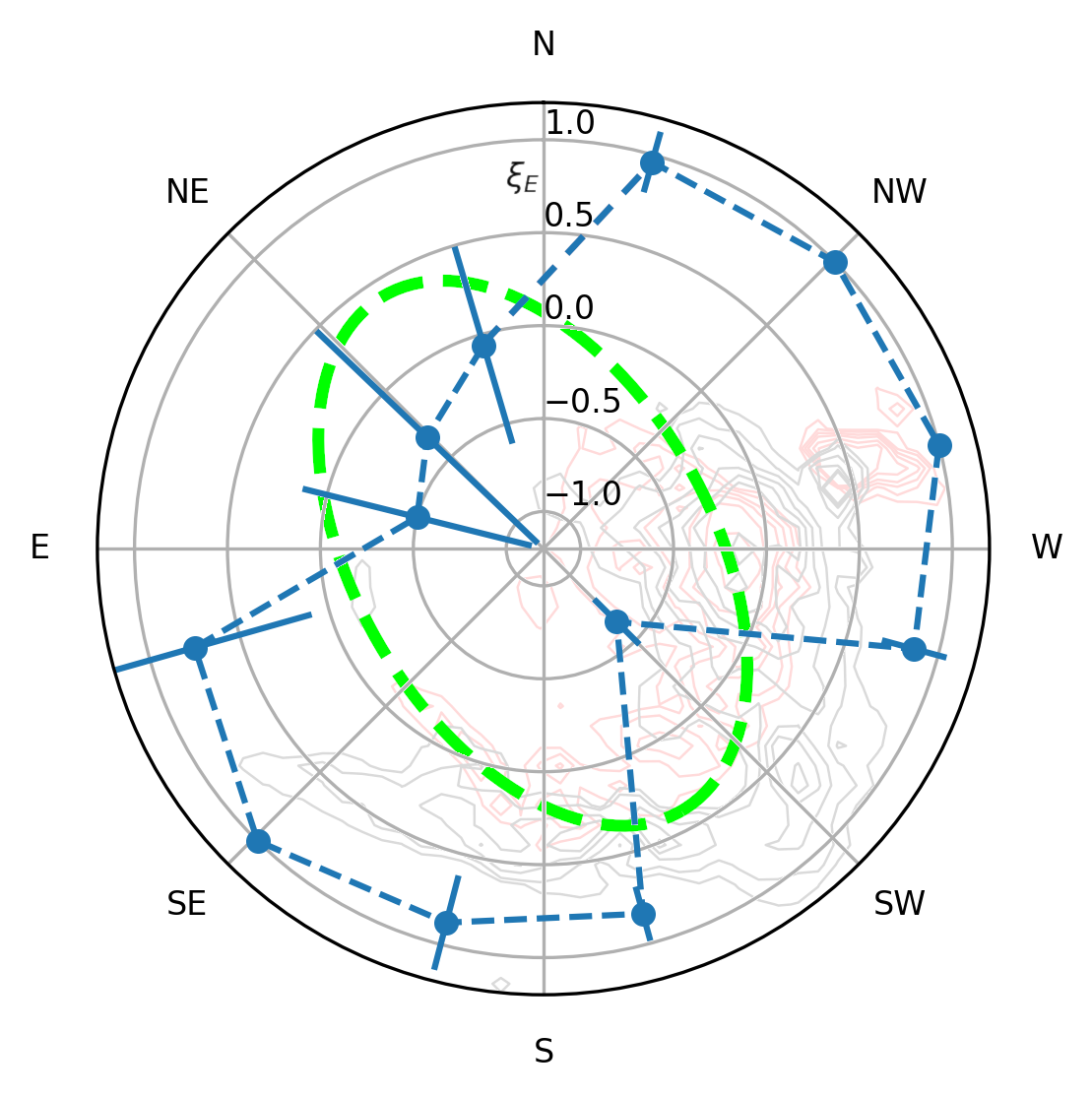}
    \caption{A radial plot of $\xi_E$. Each bin shown represents an arc of 30$^{\circ}$ of an ellipse corresponding to the boundary of HII Region D, as shown in green. $\xi_E$ was calculated using only magnetic field vectors within the Simpson bubble and PDR shown in Figure \ref{fig:spitzer}. The contours shown for the VLA C-band emission (red) and the \textit{Spitzer} 8\um\ emission (black) are identical to Figure \ref{fig:spitzer}.}
    \label{fig:circle_spit}
\end{figure}

This is supported by measuring how the magnetic field aligns to the H\textsc{ii} region itself. As both the C-band emission and the 8\um\ emission form a partial arc, we choose to model H\textsc{ii} D as an ellipse, as shown in Figure \ref{fig:spitzer}, with the same central coordinates as the Simpson bubble but at angle of 30$^\circ$ east of north. We determined the alignment between the magnetic field and the HII region structure by introducing another modification to the HRO definition,

\begin{equation}
    \phi_{E} = \frac{\textbf{B} \times \textbf{E}}{\textbf{B} \cdot \textbf{E}}
	\label{eq:phi_ell}
\end{equation}

where \textbf{E} is the local tangent to the ellipse representing the HII region boundary. Similar to $\xi_F$, this was used to calculate what will be referred to as the ellipse alignment factor ($\xi_E$), which is shown in Figure \ref{fig:circle_spit}. The $\xi_E$ value for most of the ellipse's circumference is close to 1, suggesting a strong parallel alignment of the magnetic field to the boundary of the HII region. This is as would be expected if the HII region has expanded and compressed the surrounding material \citep{arzou2021}.

However, to the north-east and south-west, the magnetic field is instead aligned preferentially perpendicular to the HII region boundary, and parallel to the major axis of the HII region itself. In the north-east, this is likely due to a low dust column density: the 850\um\ emission here is weak, with few detections of polarization. Our calculated $\xi_{E}$ values in this region have large error bars, and there is no strongly preferred alignment. However, in the south-west, 850\um\ is significantly higher, with far more detections. For this arc, it is more likely that the HII PDR has broken out of its confining shell and become `open', in a manner similar to an early champagne flow (e.g. \citealt{immer2021}). As such, the magnetic field here is tracing the early stages of a HII outflow, rather than the compression of material seen elsewhere. This is supported by the shape of the 8\um\ \textit{Spitzer} emission: while the VLA C-band emission follows a simple arc, the 8\um\ emission has an additional extended portion (see Figure \ref{fig:spitzer} that corresponds to this change in alignment.

\citet{liu2013} suggest that the expansion of Region D in particular has lead to sequential star formation across G34.26+0.15, being responsible for the formation of Regions A, B and C. The 8\um\ emission does extend sufficiently to encompass these UCH\textsc{ii} regions, supporting the suggestion there is a direct influence from H\textsc{ii} Region D. It is also possible that the 8\um\ emission around the central hub is caused by direct heating from the UCH\textsc{ii} regions themselves rather than as part of an external shock. In the hub, the magnetic field vectors are not sufficiently well-resolved to distinguish between these scenarios.

However, as discussed earlier, to the south of the Simpson bubble \citep{simpson2012} is another proposed H\textsc{ii} region, G34.24+0.13 \citep{hunter1998}, which is associated with strong 850\um\ emission and corresponds to the secondary hub. Both the 8\um\ and C-band emission appear to curve around G34.24+0.13, rather than directly overlapping it. As such, we hypothesize that the shock front caused by the larger expanding H\textsc{ii} region has intersected a pre-existing overdensity of molecular gas and dust at the current location of G34.24+0.13. This intersection may have compressed the the gas and dust, directly triggering star formation.

This hypothesis is supported by the magnetic field structure around G34.24+0.13. In Section \ref{sec:fils}, we note that Filament 6 could correspond to a secondary `hub' of the G34.26+0.15 field, exerting some gravitational influence over the local gas and dust. The magnetic field structure is largely aligned parallel to this filament, as would be expected during accretion \citep{motte2018}. However, unlike Filaments 1 and 2, this alignment is roughly constant across the entire filament, rather than becoming unaligned further from the hub. This is may be due to the field instead being aligned along the boundary of the larger H\textsc{ii} region D, suggesting that this region is of larger influence than the H\textsc{ii} region G34.24+0.13 -- it is possible even that Filament 6 was formed by a shock front, given it runs along the boundary of the 8\um\ emission. As such, Filament 6 appears to be a combination of a streamer accreting onto the secondary hub of G34.26+0.15, and an overdensity created by the expansion of H\textsc{ii} region D.

\subsection{Kinematics within G34.26+0.15}
\label{sec:kinematics}

Another factor to consider is the 3D structure of the G34.26+0.15 cloud. The observations presented here only consider the plane-of-sky structure -- i.e. the line-of-sight component is not well-constrained. In areas in which G34.26+0.15 appears to be composed of filaments with a small width ($\sim$0.2pc) this is of little concern, as the polarization vectors can to be assumed to be similarly affected by projection effects as the filaments themselves.

Some of the density structure can be inferred from the velocity analysis performed in Section \ref{sec:velocity}. There are two interpretations for the velocity gradient seen across G34.26+0.15: a large-scale flow of material from the wing component towards the hub component (or vice versa), or the presence of multiple velocity components along the line of sight. If we assume the latter, then velocity can be taken as a proxy for distance -- i.e. components moving faster are more likely to be at different distance than components moving slower. In this case, as the wing component has a 3-4 km/s difference in velocity relative to the central hub, it can be suggested that the wing containing the secondary hub is at different distance along our line-of-sight than the central hub.

\begin{figure}
	\includegraphics[width=\columnwidth]{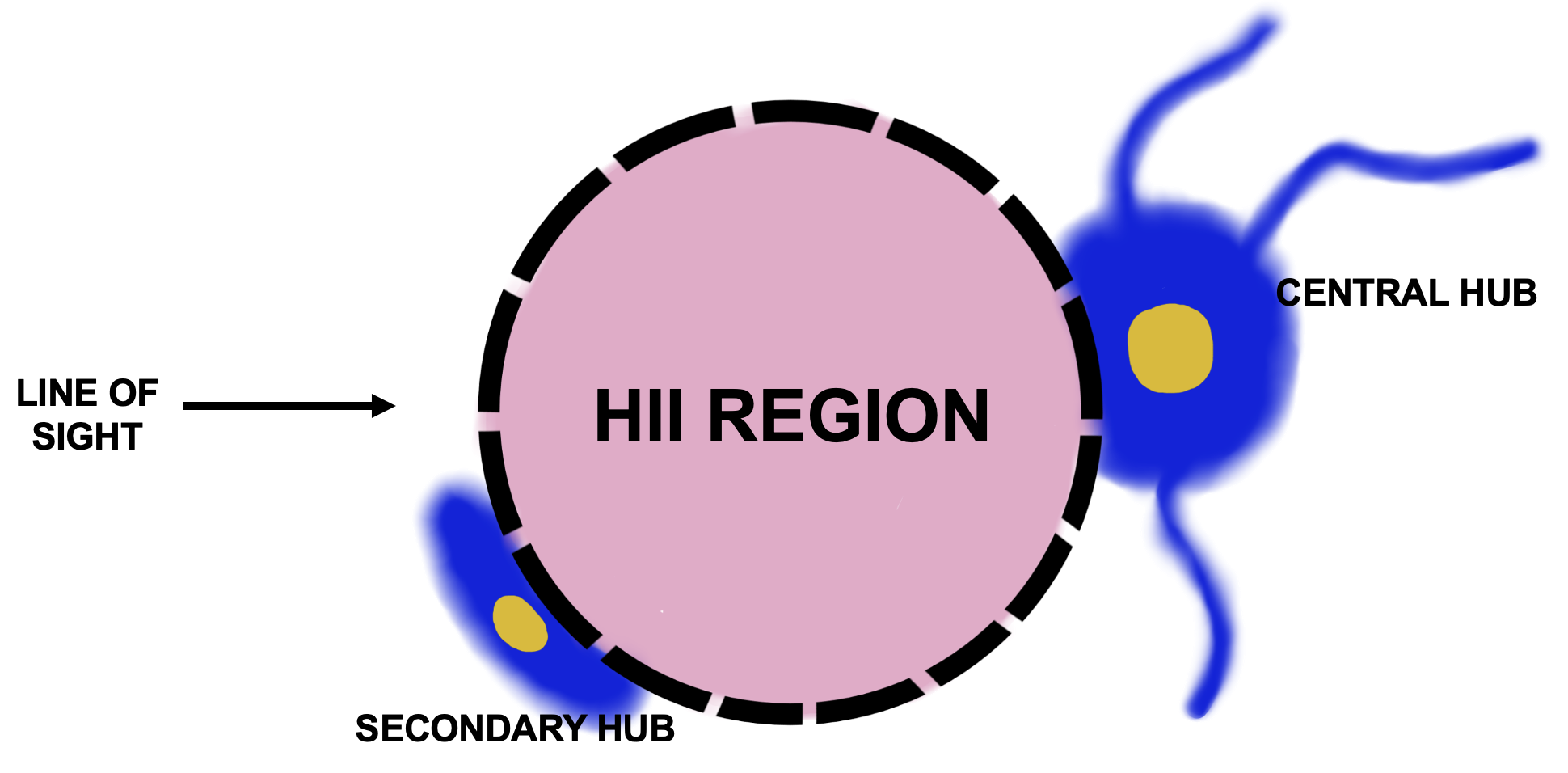}
    \caption{A diagram showing a possible configuration of G34.26+0.15. Note that although in this case the secondary hub is shown in front of the primary hub, it is equally possible that the secondary hub is behind the primary hub.}
    \label{fig:cartoon}
\end{figure}

Given that both the wing component and the central hub appear to exist on the boundary of the expanded H\textsc{ii} region D, it is unlikely that the components are entirely physically unrelated, especially if they are the result of triggered/sequential star formation. However, this is not necessarily inconsistent with a potential difference in distance, as we do not have information about the 3D structure of H\textsc{ii} region D. For example, if the H\textsc{ii} region has an approximately spherical boundary, expanding into our line of sight, as illustrated in Figure \ref{fig:cartoon}, the wing component could be located on the leading edge of the expansion, with the central hub located to the north-west. We note that in this scenario, it is unclear whether there is physical link  through a direct flow of gas. Additionally, as H\textsc{ii} Region D has largely cleared its center of dust observable at 850\um\, we can infer little about the 3D structure from the magnetic field in this region.

\section{Conclusion}
\label{sec:conc}

In this paper, we presented polarized 850\um\ observations of the molecular cloud G34.26+0.15 using the polarimeter POL-2 mounted on the SCUBA-2 camera on the JCMT.

We find that G34.26+0.15 displays markedly different magnetic field structure across its two distinct halves. The north-west region has an average angle of 103.5$^\circ$ and the south-east region has an average angle of 57.7$^\circ$ (east of north). We propose that this discrepancy is due to the reshaping of the cloud by an expanding H\textsc{ii} region, located to the west of the central bright peak, which has affected the magnetic field structure of the south-east region. Additionally, we used the DCF method to estimate the magnetic field strength across each components -- we find a higher field strength of 1.39 mG for the south-eastern wing region compared to 1.12 mG for Filament 1 in the north-west, although these are subject to large uncertainties inherent in DCF, especially the lack of 3D morphological information.

We also investigated the magnetic field in relation to the filamentary structure of G34.26+0.15. Here, we find that the field in the north-west half displays a strong parallel alignment to the filaments at high densities, with weaker alignment at low densities, as would be expected in a gravity-dominated hub-filament system \citep{motte2018}. However, the magnetic field in the south-east half instead appears to align to the boundary of the H\textsc{ii} region D. This is suggested to be indicative of potential sequential star formation within G34.26+0.15, supporting the proposed sequence as suggested by \citet{liu2013} and \citet{xu2016}.

Despite appearing to be a `typical' hub-filament system when observed with 850\um\ Stokes I alone, the inferred magnetic field data presented shows a more complex arrangement, with G34.26+0.15 being vastly altered by strong H\textsc{ii} region feedback. Given expanding H\textsc{ii} regions exist ubiquitously within high-mass star forming systems, this would suggest such massive reshaping events are common in hub-filament systems, potentially altering star formation rates. This also demonstrates the value of mapping magnetic fields, even in gravity-dominated areas, as a method of investigating stellar feedback effects.

\section*{Acknowledgements}

K.P. is a Royal Society University Research Fellow, supported by grant number URF\textbackslash R1\textbackslash 211322.

The James Clerk Maxwell Telescope is operated by the East Asian Observatory on behalf of The National Astronomical Observatory of Japan; Academia Sinica Institute of Astronomy and Astrophysics; the Korea Astronomy and Space Science Institute; the National Astronomical Research Institute of Thailand; Center for Astronomical Mega-Science (as well as the National Key R$\&$D Program of China with No. 2017YFA0402700). Additional funding support is provided by the Science and Technology Facilities Council of the United Kingdom and participating universities and organizations in the United Kingdom, Canada and Ireland. Additional funds for the construction of SCUBA-2 were provided by the Canada Foundation for Innovation.

The data used in this paper were taken under the project code M16AD003.  The authors thank Paul Ho for his award of JCMT Director's Discretionary Time (DDT) to this project.

The Starlink software (Currie et al. 2014) is supported by the East Asian Observatory.  This work made use of Astropy:\footnote{http://www.astropy.org} a community-developed core Python package and an ecosystem of tools and resources for astronomy \citep{astropy:2013, astropy:2018, astropy:2022}.

The authors wish to recognize and acknowledge the very significant cultural role and reverence that the summit of Maunakea has always had within the indigenous Hawaiian community.  We are most fortunate to have the opportunity to conduct observations from this mountain.

\section*{Data Availability}

The data used in this paper are available in the JCMT archive hosted by the Canadian Astronomy Data Centre (CADC) under the project code M16AD003. The reduced data from this project can found from the CADC at [DOI]. Filamentary data of G34.26+0.15 are available upon request. 



\bibliographystyle{mnras}




\appendix

\section{$^{13}$CO J=$3\to2$ Data}
\label{sec:13co}

\begin{figure*}
\begin{minipage}[ht]{\textwidth}
    \includegraphics[width=\textwidth]{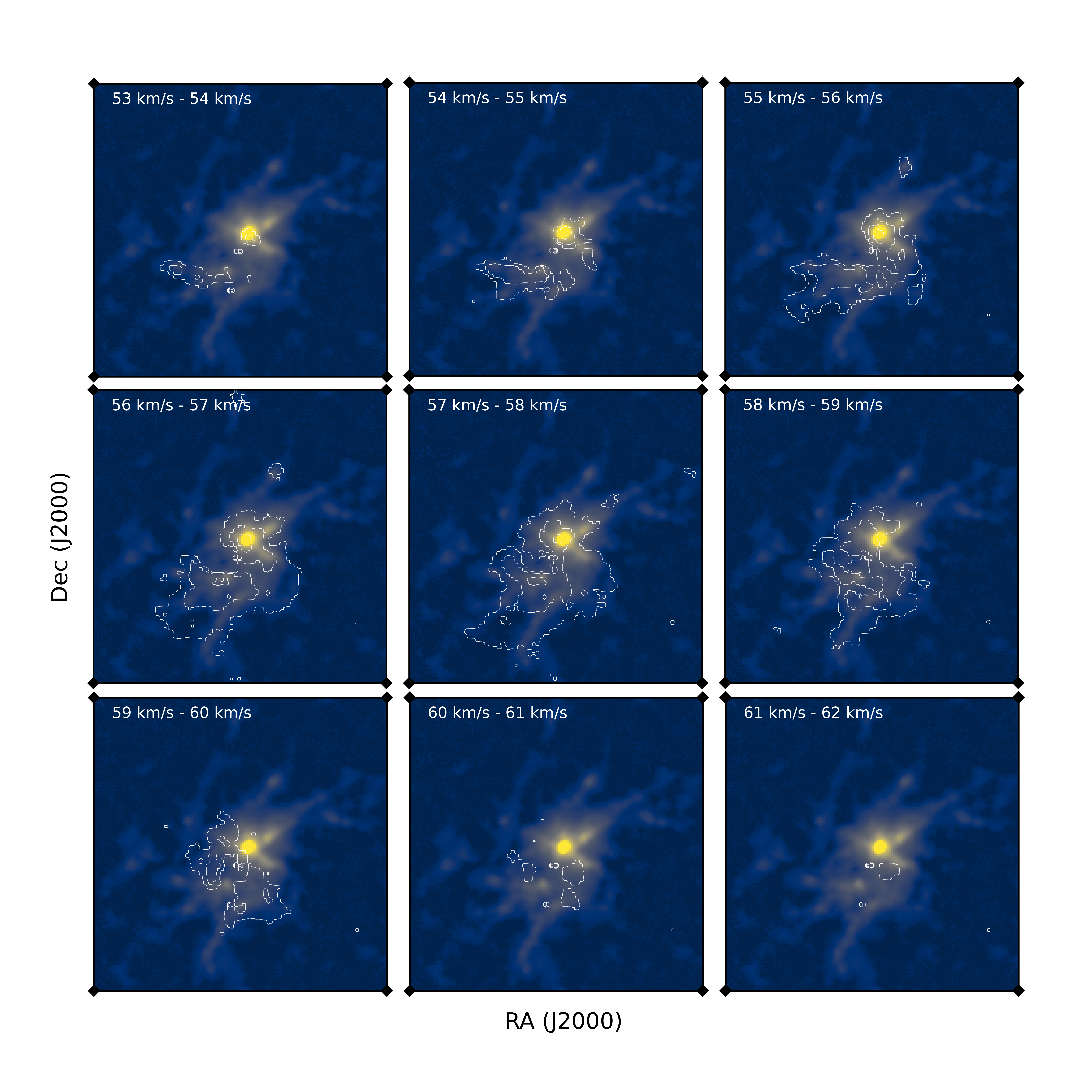}
    \vspace{-40pt}
    \caption{$^{13}$CO J=$3\to2$ channel map contours overlaid on the 850\um\ Stokes I map of G34.26+0.15. Channel widths are 1km/s. For details about this data, see Section \ref{sec:velocity}}
    \label{fig:vels_all_13co}
\end{minipage}
\end{figure*}

The HARP data from the CHIMPS survey \citep{rigby2016} includes both $^{13}$CO J=$3\to2$ and C$^{18}$O J=$3\to2$ line data. For the DCF analysis presented in this paper (see Section \ref{sec:velocity} only the C$^{18}$O J=$3\to2$ was used, due to the $^{13}$CO J=$3\to2$ presenting self-absorption effects at high intensities. The $^{13}$CO J=$3\to2$ data is presented in Figure \ref{fig:vels_all_13co}.

\bsp	
\label{lastpage}
\end{document}